\documentclass[lettersize,journal]{IEEEtran}
\IEEEoverridecommandlockouts
\usepackage{cite}
\usepackage{amsmath,amssymb,amsthm,amsfonts}
\usepackage{algorithm}
\usepackage{algpseudocode}
\usepackage{graphicx}
\usepackage{subfigure}
\usepackage{textcomp}
\usepackage{xcolor}
\usepackage{stfloats}
\usepackage{mathtools}
\def\BibTeX{{\rm B\kern-.05em{\sc i\kern-.025em b}\kern-.08em
    T\kern-.1667em\lower.7ex\hbox{E}\kern-.125emX}}
\graphicspath{{Figures/}{./}}
\usepackage[numbers,sort&compress]{natbib}

\renewcommand{\baselinestretch}{0.945}

\usepackage{array}
\usepackage{verbatim}
\usepackage{diagbox}
\usepackage{makecell}
\usepackage{multirow}
\usepackage{bm}

\usepackage{pifont}
\newcommand{\cmark}{\color{green} \ding{51}}%
\newcommand{\xmark}{\color{red} \ding{55}}%

\newtheorem{theorem}{Theorem}
\newtheorem{lemma}[theorem]{Lemma}

\begin{document}

\algnewcommand\algorithmicswitch{\textbf{switch}}
\algnewcommand\algorithmiccase{\textbf{case}}
\algnewcommand\algorithmicassert{\texttt{assert}}
\algnewcommand\Assert[1]{\State \algorithmicassert(#1)}%
\algdef{SE}[SWITCH]{Switch}{EndSwitch}[1]{\algorithmicswitch\ #1\ \algorithmicdo}{\algorithmicend\ \algorithmicswitch}%
\algdef{SE}[CASE]{Case}{EndCase}[1]{\algorithmiccase\ #1}{\algorithmicend\ \algorithmiccase}%
\algtext*{EndCase}%

\title{Cell-Free Massive MIMO Under Mobility: \\ A Fairness-Differentiated Handover Scheme

\thanks{The authors are with the Chair of Mobile Communications and Computing at RWTH Aachen University, Aachen 52072, Germany (email: \{yunlu.xiao, petrova, simic\}@mcc.rwth-aachen.de). \\ The preliminary conference version of this work \cite{wcnc} was presented at IEEE WCNC 2024, Dubai.}}

%


\author{\IEEEauthorblockN{Yunlu~Xiao, Marina Petrova,
and
Ljiljana~Simić}}

\maketitle

\begin{abstract}
While cell-free massive MIMO (\mbox{CF-mMIMO}) offers high and uniform network-wide throughput in static networks, its performance in mobile networks is not yet fully addressed. In this paper, we evaluate the throughput performance of urban mobile CF-mMIMO networks under a comprehensive throughput model and show that it suffers from large performance degradation due to the combined effect of channel aging and handover overheads. To restore the uniformly good performance of \mbox{CF-mMIMO} under mobility, we formulate a novel optimization problem to maximize the nett throughput that considers both channel aging and handover cost. We derive a \mbox{near-optimal} solution \textit{nearOpt} for our transformed and relaxed optimization problem with Newton's method. We then design a heuristic handover algorithm, \textit{FairDiff}, to differentiate prioritized and optional handovers using a policy threshold based on Jain's fairness index, in order to achieve uniform throughput over the network. Our extensive evaluation of the mobile throughput performance of our handover schemes in realistic urban mobile networks shows that, unlike the existing literature benchmarks that obtain very low throughput under mobility, our \textit{FairDiff} scheme consistently achieves the near-optimal throughput comparable to \textit{nearOpt} and highest network-wide throughput with the lowest computational complexity among all considered schemes. We thus for the first time propose a handover scheme that delivers the promise of uniformly good throughput for mobile CF-mMIMO, making it a feasible architecture for practical mobile networks.\looseness=-1
\end{abstract}


\vspace{-0.1cm}
\section{Introduction}

\IEEEPARstart{C}{ell}-free massive multiple-input-multiple-output (\mbox{CF-mMIMO}) has been proposed \cite{cfvs} as a disruptive architecture for 6G mobile networks \cite{CFmag2026}. Unlike the traditional cellular network, where one base station only serves UEs within its cell, CF-mMIMO coordinates all access points (APs) in the network to serve a given user (UE). This architecture thereby combines the high throughput benefit of massive MIMO and the uniform throughput benefit of distributed MIMO \cite{MIMObook}. With no cell boundaries nor changes in serving APs, the CF-mMIMO network thus eliminates both the \mbox{cell-edge} effect for the \mbox{poorly-served} UEs and the challenge of handover. CF-mMIMO thereby provides significantly \textit{higher} and more \textit{uniform} through over the network, promising to deliver ubiquitous 6G connectivity with ``uniformly great service for everyone'' \cite{CFmag2026, CFfirst}. However, in practical networks, coordinating all APs to simultaneously serve all UEs brings very high signal processing and signaling overheads. Scalable CF-mMIMO is thus proposed \cite{cfsim} to limit the system cost by managing the APs via several disjoint CPU clusters and serving a given UE with a finite cluster set. This \mbox{UE-centric} architecture effectively forms a serving AP set that surrounds the UE, and thus reduces the cluster-edge effect and achieves both high and uniform throughput for all UEs in a static network \cite{scalableCF}. \looseness=-1
  

However, in case of UE mobility, the performance of the scalable CF-mMIMO architecture is degraded by both channel aging and handover. Firstly, the serving AP set selection and signal processing of CF-mMIMO depend on channel state information, which is quickly outdated due to a fast-changing mobile channel, i.e., the channel aging effect \cite{aging}. Secondly, maintaining a UE-centric serving AP set for a mobile UE requires frequent changes of the serving APs, leading to throughput degradation due to high handover interruption time cost \cite{xiao}. This raises the question of whether CF-mMIMO would maintain its uniformly good throughput performance under mobility. \looseness=-1

Although the performance of mobile CF-mMIMO is impacted by both channel aging and handover, the prior literature either only optimizes the throughput under channel aging while entirely neglecting handover analysis \cite{aging, scalableaging, agingPredict, agingOptimize, drlJP}, or solely reduces the handover rate \cite{pomdp, SoftHOcf, beerten2025}. In \cite{myGCpaper} we showed that such incomplete mobility models bias the conclusions drawn in \cite{agingPredict, agingOptimize, drlJP, pomdp, SoftHOcf, beerten2025}. Specifically, \cite{agingPredict, agingOptimize, drlJP} optimize the serving set to achieve high throughput under channel aging, but the potentially high handover rate may still degrade the throughput under mobility. Only a handful of works have studied handover schemes for CF-mMIMO \cite{pomdp, SoftHOcf, beerten2025}. However, these prior schemes reduce the handover rate by indiscriminately suppressing handovers for all UEs, neglecting the channel aging impact of an outdated serving set on the throughput. This has a particularly detrimental impact on the \mbox{poorly-served}, e.g. $95^\text{th}$ percentile UEs, reintroducing the \mbox{edge-effect} and undermining the uniform-performance design goal of CF-mMIMO. In this paper, we re-evaluate the benchmark schemes \cite{pomdp, SoftHOcf, beerten2025} using a comprehensive throughput model that considers both channel aging and handover interruption time costs. We show that they in fact do not provide uniformly good throughput performance across the network, due to indiscriminately reducing the UE handover rate of \mbox{CF-mMIMO} by applying the same handover policy to all UEs. Namely, the handover schemes from the prior literature fail to deliver the promise of \mbox{CF-mMIMO} for uniformly good throughput performance under mobility. \looseness=-1

To address this gap, we formulate a novel optimization problem to maximize the nett throughput under both channel aging and handover and derive a near-optimal solution, \textit{nearOpt}, for the relaxed and transformed problem. We then design a novel \mbox{fairness-differentiated} handover scheme, \textit{FairDiff}, aiming to preserve the premise of uniformly good performance for all UEs in \mbox{CF-mMIMO} by employing Jain's fairness index to quantify the uniformity of throughput over the network and differentiate the handover policy of UEs based on this. As the heuristic solution of the optimization problem, \textit{FairDiff} achieves equivalently good performance as \textit{nearOpt}, but with an order of magnitude lower complexity.\looseness=-1



Our key contributions are:

\begin{itemize}

\item We present the first handover analysis of mobile \mbox{CF-mMIMO} using a comprehensive throughput model that considers both channel aging and handover overheads. We show that under this model, the existing benchmark handover schemes perform poorly due to solely reducing the handover rate of all UEs indiscriminately or solely maximizing the throughput under channel aging. They thus obtain low and \mbox{non-uniform} throughput under mobility, in contrast to the design goal of CF-mMIMO.

\item We formulate a novel optimization problem that maximizes the nett throughput taking into account both channel aging and handover overheads and derive a near-optimal solution \textit{nearOpt} by directly solving the relaxed optimization problem using Newton's method.

\item We propose a fairness-differentiated handover algorithm \textit{FairDiff} as a \mbox{low-complexity} heuristic solution of our optimization problem. We introduce Jain's fairness index to quantify the uniformity of throughput over all UEs and design a handover policy threshold based on it to differentiate between: necessary handovers to be prioritized for maintaining high throughput and optional handovers to be minimized for reducing handover overheads.

\item We show that both our \textit{nearOpt} and \textit{FairDiff} schemes consistently and significantly outperform the benchmark handover schemes, especially for the \mbox{poorly-served} UEs, via an extensive throughput performance evaluation under realistic urban mobility. Importantly, our low-complexity \textit{FairDiff} handover scheme for the first time delivers in practice the promise of \mbox{CF-mMIMO} of uniformly good performance, for not only static but also mobile networks.

\end{itemize}




The rest of the paper is organized as follows: Sec. \ref{relate} details the related work on mobile CF-mMIMO, Sec. \ref{model} presents our system model, Sec. \ref{optProblem} presents the formulation and solution of our novel optimization problem, Sec. \ref{algo} proposes our \textit{FairDiff} handover algorithm, \mbox{Sec. \ref{results}} presents our results, and \mbox{Sec. \ref{conclude}} concludes the paper. \looseness=-1


\section{Related Work}
\label{relate}

The impact of channel aging and handover on the performance of mobile \mbox{CF-mMIMO} has largely been studied separately in the prior literature. Channel aging is first modelled in \cite{aging} to present the throughput degradation of the original unscalable \mbox{CF-mMIMO} under mobility. Channel aging for scalable \mbox{UE-centric} \mbox{CF-mMIMO} is furthermore studied in \cite{scalableaging}, showing its great impact on throughput under high mobility. To improve the throughput under channel aging, a finite impulse response Wiener predictor is proposed in \cite{agingPredict} and the duration of the communication block is optimized in \cite{agingOptimize}. In \cite{xiao}, we first analyzed the mobility impact of handover in terms of a delay cost, and showed that without considering channel aging, the \mbox{UE-centric} \mbox{CF-mMIMO} architecture can achieve high and uniform throughput for moderate mobility levels without a selective handover scheme. However, we showed in \cite{myGCpaper} that when \textit{both} channel aging and handover are considered, the combined mobility impact brings significant throughput degradation to UE-centric CF-mMIMO, in fact rendering it unattractive for practical mobile networks without a smart handover strategy in place. Crucially, this means that capturing the throughput performance of mobile CF-mMIMO requires modeling both channel aging and handover interruption time. \looseness=-1

\begin{table}[t]
\caption{Related Work on Mobile CF-mMIMO.}
		\centering
		\footnotesize
		\setlength{\tabcolsep}{3.2pt}
        \renewcommand{\arraystretch}{1}
		\begin{tabular}{l c c c c} 
			{Reference}  &  {HO} & {HO} & {Channel} &  {Realistic} \\ 
			{studies}    & {scheme} & {cost}   & {aging}          & {topology/mobility} \\ 				
    \hline
    \hline
	\cite{aging}  	 & \xmark  & \xmark   & \cmark & \xmark\; \color{black}/ \color{red} \xmark \\
    	\cite{xiao}  	& \xmark  & \cmark  & \xmark  & \xmark \; \color{black}/ \color{red} \xmark \\
	\cite{scalableaging, agingPredict, agingOptimize}  	& \xmark & \xmark &  \cmark   & \xmark \; \color{black}/ \color{red} \xmark  \\	

	\cite{myGCpaper}  & \xmark & \cmark  & \cmark  & \xmark \; \color{black}/ \color{red} \xmark  \\	
    \cite{drlJP} & \xmark & \xmark  & \cmark  & \cmark \; \color{black}/ \color{green} \cmark  \\
	\cite{pomdp}  & POMDP & \xmark  & \cmark  & \xmark \; \color{black}/ \color{red} \xmark \\	
    \cite{SoftHOcf} & UE-performance-aware & \xmark  & \cmark  & \cmark \; \color{black}/ \color{red} \xmark \\
	\cite{beerten2025} & Hysteresis & \xmark  & \xmark  & \xmark \; \color{black}/ \color{red} \xmark \\	


	\textbf{This paper}		& \textbf{\textit{FairDiff (+ nearOpt)}}  & \cmark & \cmark  & \cmark \; \color{black}/ \color{green} \cmark \\ 
		\end{tabular}

\label{related_work}
\end{table}
 
The change of serving AP set and the related handover decision problem of CF-mMIMO have been rarely studied. In \cite{drlJP} a deep learning-based serving AP set selection for mobile \mbox{CF-mMIMO} is proposed to optimize the throughput and reduce the signaling during the mobility period. However, it neglects entirely the handover cost of maintaining the optimized serving set. A UE-performance-aware handover scheme is proposed in \cite{SoftHOcf} to reduce the handover rate by only triggering the handover of a given UE when its performance has fallen by a certain margin. The hysteresis scheme in \cite{beerten2025} further reduces the handover rate by additionally requiring that the performance after handover is better than before handover by a certain margin. A scheme to reduce the handover rate for an optimized target throughput is derived in \cite{pomdp} by modeling the handover procedure as a partially observable Markov decision process (POMDP) problem. However, these prior benchmarks solely focus on reducing the handover rate, while entirely neglecting the impact of handover on the throughput performance. By contrast, our work derives the handover strategy by directly maximizing the throughput under both channel aging and handover overhead. Using this comprehensive throughput model under mobility, we thus show that our proposed handover schemes significantly outperform all prior benchmarks \cite{pomdp, SoftHOcf, beerten2025}. \looseness=-1

Finally, we note that the vast majority of prior literature on mobile CF-mMIMO \cite{aging, xiao, scalableaging, agingPredict, agingOptimize, pomdp, beerten2025} has assumed a random uniform network topology and random way point (RWP) mobility pattern. A realistic topology and mobility pattern of the area of Tokyo are assumed in \cite{drlJP} to show the highly non-uniform urban network topology cannot be well modelled by the widely assumed random uniform distribution and RWP mobility. However, no handover analysis is included in this work. A realistic urban topology of the city area of Seoul is considered in \cite{SoftHOcf}, but with RWP UE mobility. By contrast, we evaluate our handover schemes not only in the scenario of randomly uniformly distributed APs and RWP UE mobility, but also in realistic urban network distributions and pedestrian/vehicular mobility patterns of two city areas.


\section{System Model}
\label{model}
\subsection{Network Architecture \& Mobility Model}
\label{Arch}

\begin{figure*}[!tb] 
	\centering
	\subfigure[PPP, example RWP track]{
		\label{city.sub.1}
		\includegraphics[width=0.18\linewidth]{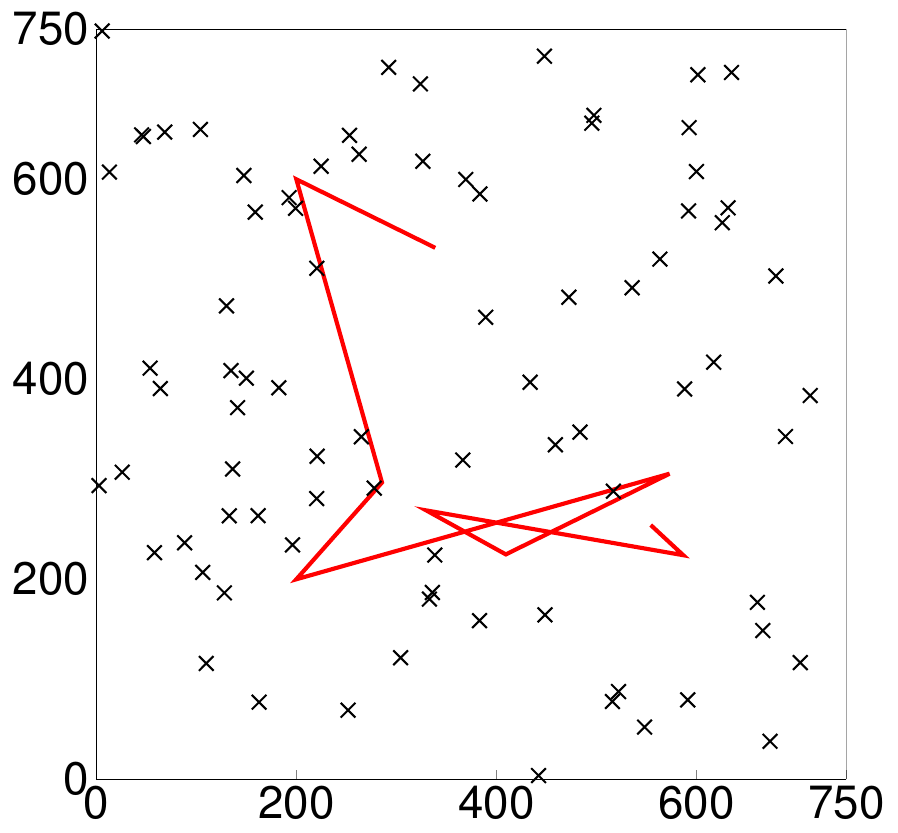}}
	\subfigure[Seoul, 50 pedestrian tracks]{
		\label{city.sub.2}
		\includegraphics[width=0.18\linewidth]{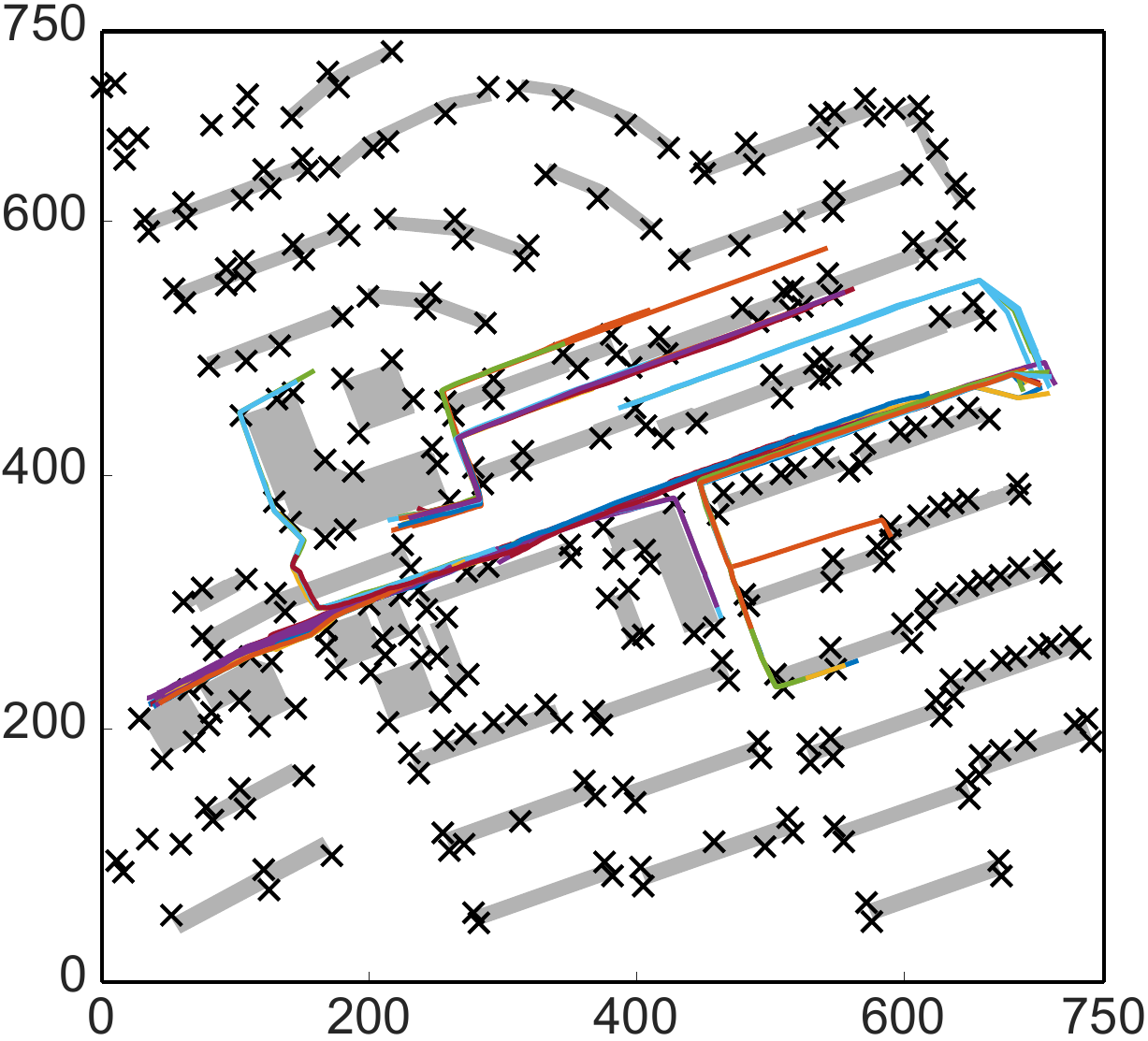}}
	\subfigure[Seoul, 50 vehicular tracks]{
		\label{city.sub.4}
		\includegraphics[width=0.18\linewidth]{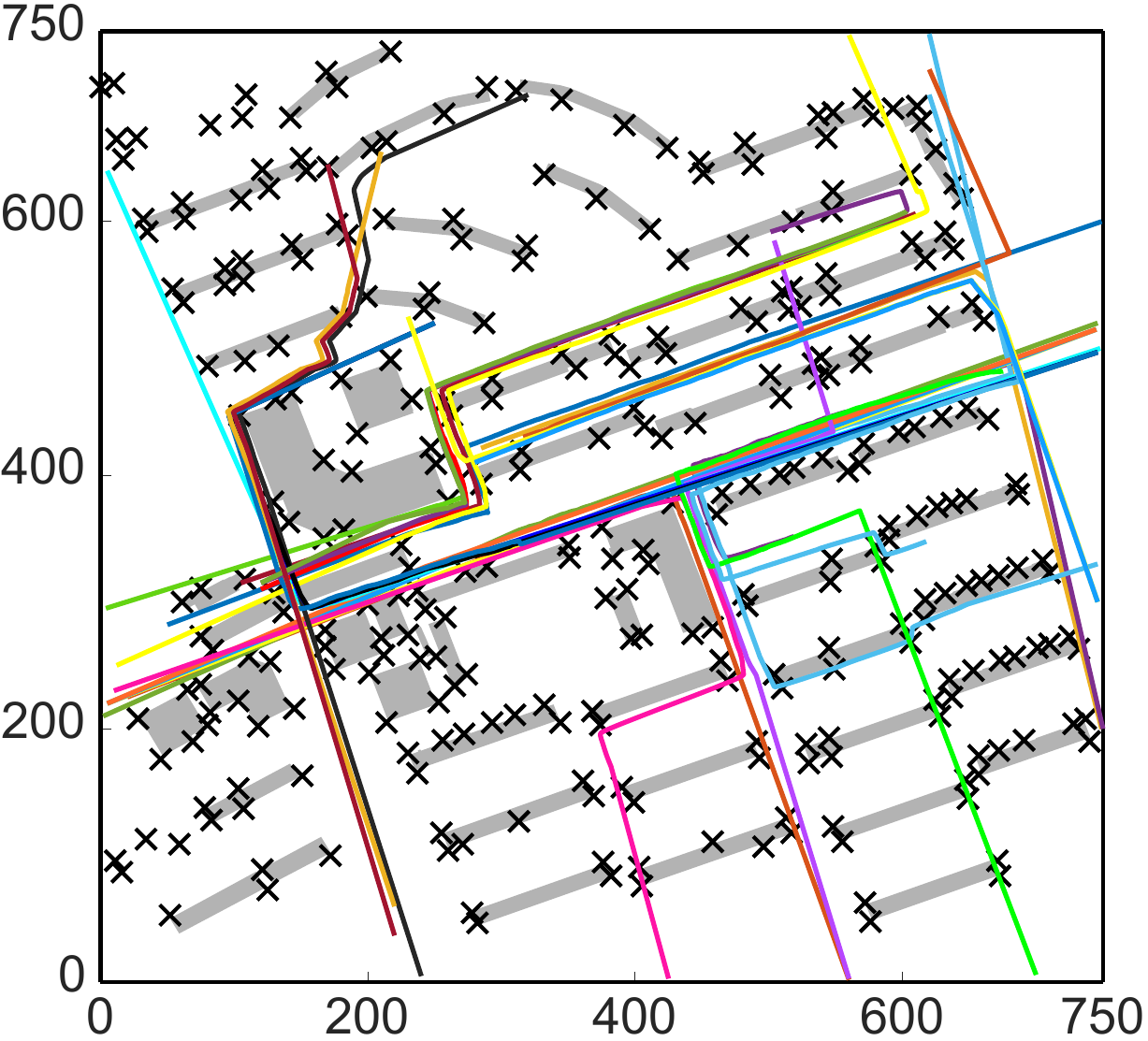}}
	\subfigure[Frankfurt, 50 pedestrian tracks]{
		\label{city.sub.3}
		\includegraphics[width=0.18\linewidth]{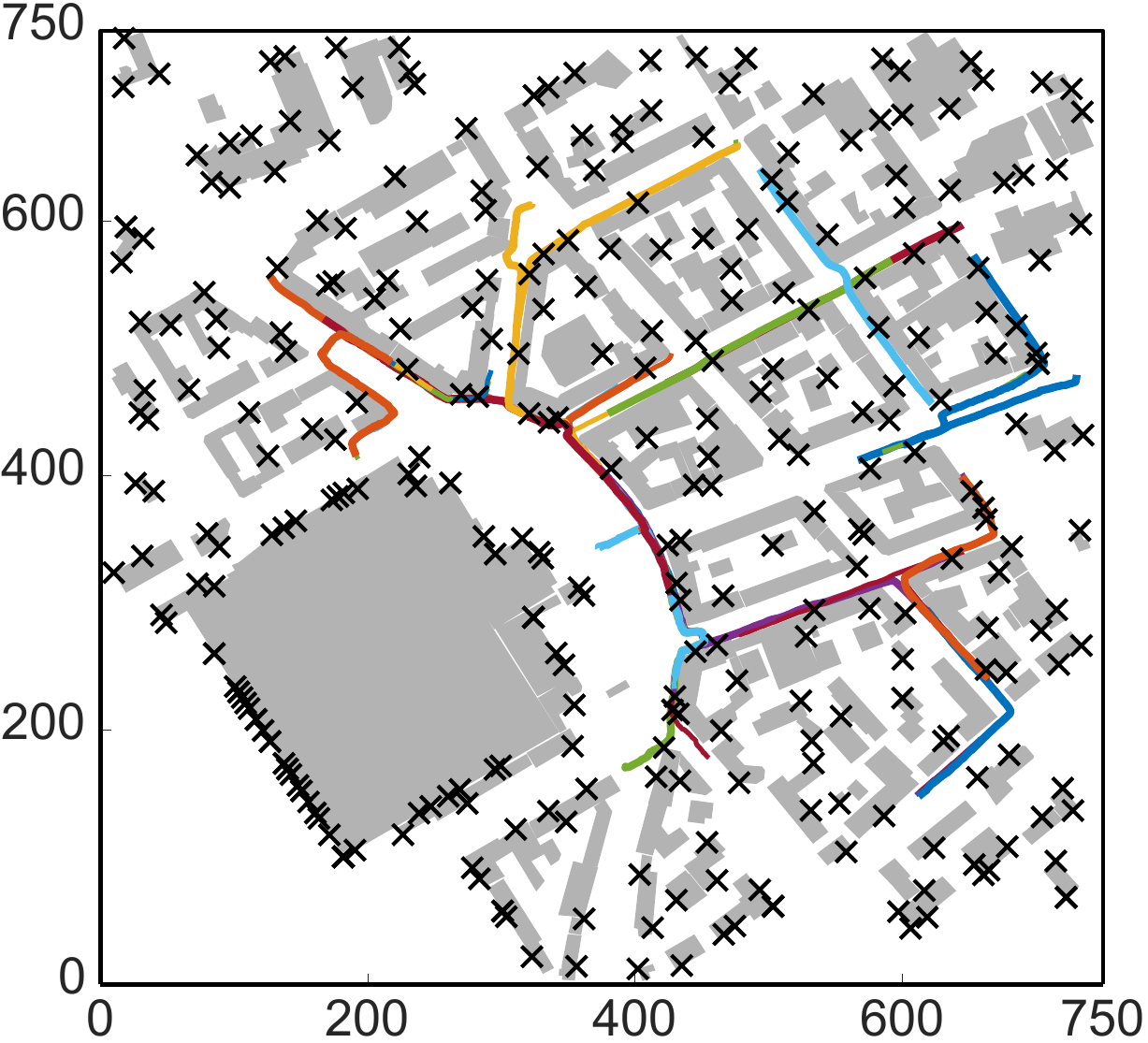}}
	\subfigure[Frankfurt, 50 vehicular tracks]{
		\label{city.sub.5}
		\includegraphics[width=0.18\linewidth]{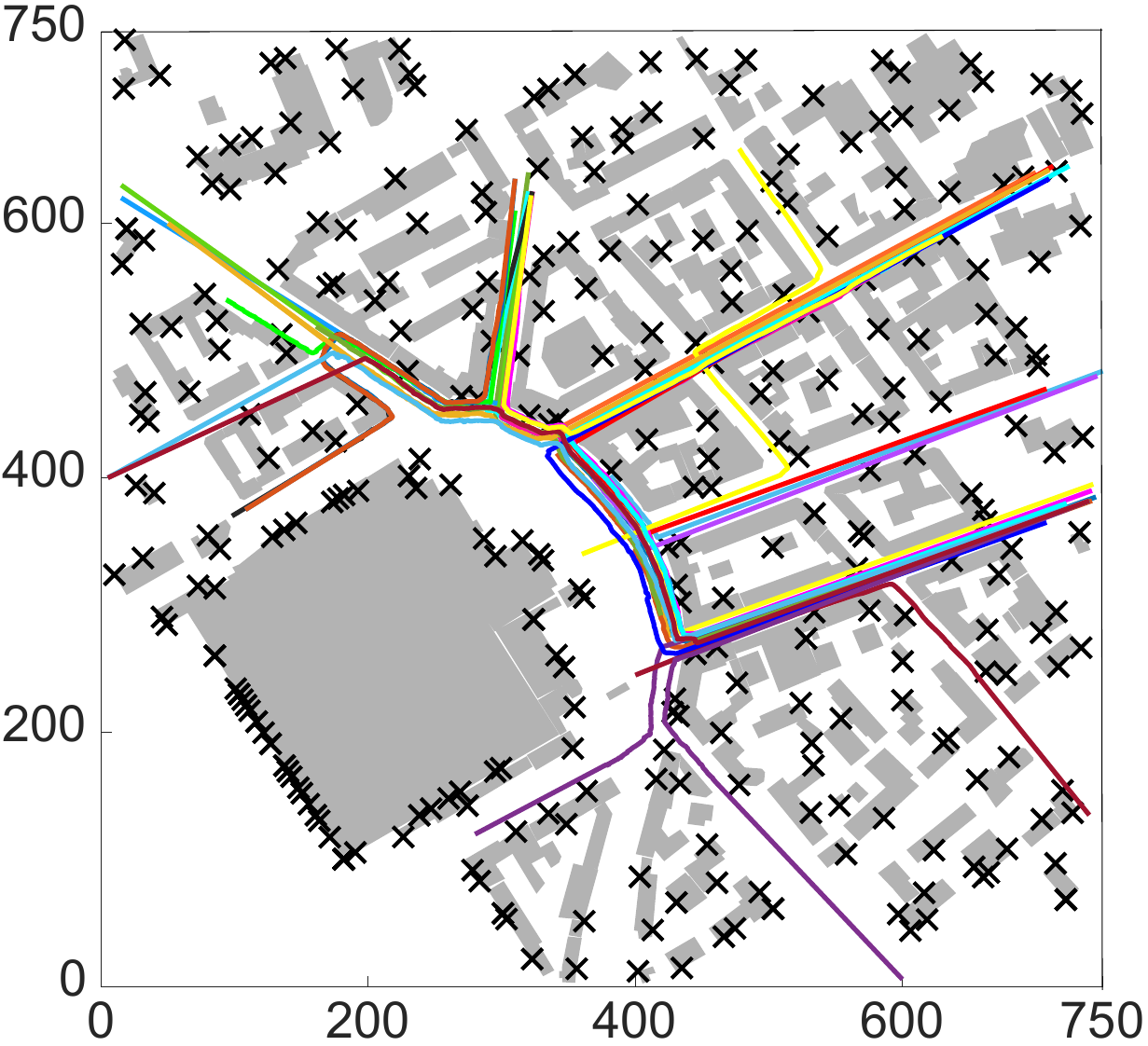}}
	\caption{Network topologies and UE mobility patterns, with coordinates in meters, buildings shown as gray polygons, AP locations marked with black crosses, and UE tracks marked with colored lines (not all tracks individually visible due to overlap).}
	\label{city}
\end{figure*}

We consider a network with \mbox{$M=308$} static APs with antenna height of 10~m and \mbox{$K=50$} mobile UEs with antenna height of 1~m distributed in an \mbox{$S=750 \text{ m} \times 750 \text{ m}$} area, resulting in an AP density of \mbox{$\lambda = 547 \text{ AP/km}^2$} and a UE density of $\lambda_u = 89 \text{ UE/km}^2$, corresponding to a \mbox{typical-density} urban network \cite{poster} and satisfying the massive MIMO configuration of $M \gg K$ for sufficient spatial diversity and channel hardening \cite{MIMObook}. We consider a PPP network with $M$ APs randomly uniformly distributed in the network. The $K$ UEs are initially distributed according to a PPP distribution and move according to the RWP model, with a random direction, \mbox{Rayleigh-distributed} transition lengths, and speeds of $v=1$~m/s and 10~m/s, corresponding to pedestrian and vehicular mobility, respectively. We show in Fig. \ref{city.sub.1} an example RWP UE track in the PPP network. To study realistic urban mobility, we also consider two real city areas in Seoul and Frankfurt with qualitatively dissimilar layouts as our study areas. We obtain the realistic AP distribution by placing the APs at edges or corners of the real buildings in these two areas, obtained via publicly available databases \cite{shp}. We then generate pedestrian tracks of all UEs with an average speed of $v=1$~m/s via Viswalk \cite{viswalk} and vehicular tracks of all UEs with $v=10$~m/s via Vissim \cite{vissim}, representing typical urban mobility. For all considered scenarios, none of the UEs ever move out of area $S$ and thus the UE density does not change during the entire simulation period. Fig. \ref{city} shows the considered AP distributions and UE tracks. 

\begin{figure}[t] 
	\centering{
		\includegraphics[width=0.9\linewidth]{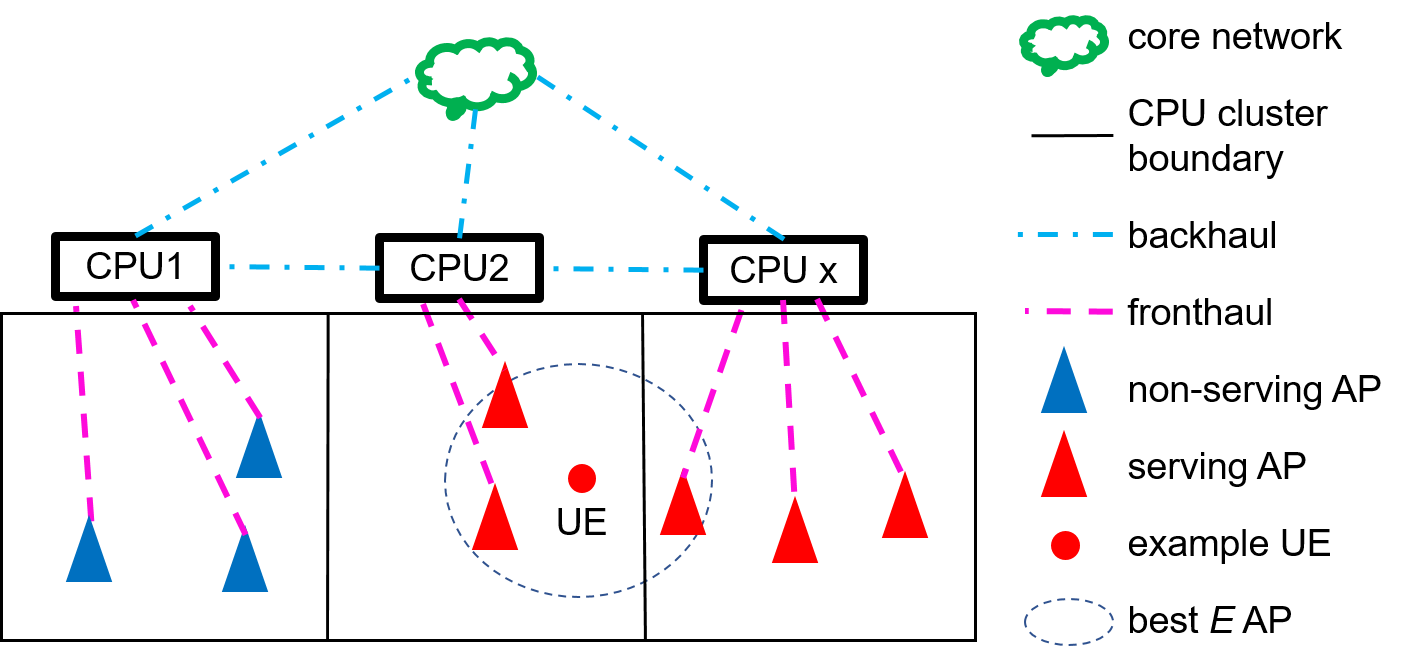}}
	\caption{Illustration of scalable CF-mMIMO architecture.}
	\label{net}
\end{figure}

We consider the scalable UE-centric CF-mMIMO architecture illustrated in Fig. \ref{net}, compatible with the O-RAN architecture in \cite{ORAN2025, LiORAN}. The APs are assigned to disjoint square CPU clusters and connected to the CPU of the cluster via fronthaul. We denote the average number of APs in one CPU cluster, i.e., the CPU cluster size, as $Q$. The CPUs connect to each other and eventually to the core network via backhaul. To avoid unlimited signaling and precoding complexity, each UE is served by a subset of APs, which can be managed by multiple CPUs. We adopt the dynamic cooperation matrix in \cite{scalableCF} to describe the connections between all APs and UEs over the network. We define the $M \times K$ cooperation matrix at time $t$ as $D[t]$: at time instant $t$, if there is a connection between AP $m$ and UE $k$, $D_{mk} [t] = 1$, otherwise $D_{mk} [t] = 0$. We define the average serving set size of all UEs as $G =(\sum_{k=1}^{K} \sum_{m=1}^{M} D_{mk}[t])/K $.

For the sake of simplicity and to make the derivation of a handover algorithm in Sec. \ref{optSolve} tractable, we assume the following simple AP selection scheme\footnote{We will additionally study in Sec. \ref{algoAna:FairDiff} the performance of our proposed \textit{FairDiff} handover algorithm with other representative AP selection schemes to verify its good performance in general mobile CF-mMIMO networks} representative of UE-centric \mbox{CF-mMIMO} literature \cite{cfsim, xiao, myGCpaper, GMM}: the example UE first selects $E$ APs with the best channel condition, and is then served by all APs in the CPU clusters that these $E$ APs belong to.\looseness=-1

\subsection{Channel \& Signal Model}
\label{signal}

We assume the Hata-COST231 \mbox{three-slope} \mbox{log-distance}\footnote{Propagation in the realistic urban networks assumed in Sec. \ref{Arch} is also in practice site-specific with building shadowing and street canyon effect, which influences the channel quality of candidate AP sets \cite{mywcnc25} but nonetheless affects all handover algorithms equally. Since our focus here is on handover design, we defer the study under realistic propagation to future work.} path loss channel model \cite{cfvs}, which gives the path loss with shadowing between AP $m$ and UE $k$ as $L_{m k}[t] = \overline{\mathrm{L}}_{m k}[t] 10^{\sigma z_{mk} /10}$, with $\sigma$ as the shadowing deviation, $z_{mk} \sim N(0,1)$, and the average path loss $\overline{\mathrm{L}}_{m k}$ given by (52) in \cite{cfvs}. \looseness=-1

The AP-UE channel is assumed to experience Rayleigh fading. The channel between AP $m$ and UE $k$ at time $t$ is modelled as $h_{mk}[t] \sim N_{C}(0,R_{mk}[t])$, where $R_{mk}$ represents the large-scale fading of the link between AP $m$ and UE $k$ \cite{scalableCF}, given by \mbox{$R_{mk}[t] = 1/L_{mk}[t]$}. The average \mbox{signal-to-noise} ratio (SNR) of the channel between AP $m$ and UE $k$ is given by \mbox{$\beta_{mk}[t]=p_{mk}/(L_{mk}[t]n_0)$}, where $p_{mk}$ is the transmit power and $n_0$ is the Gaussian noise between AP $m$ and UE $k$ over a bandwith of 20 MHz. The SNR is used to represent the channel quality during serving AP selection and handover decisions, since the throughput performance is correlated with the SNR \cite{aging}. \looseness=-1

\begin{figure}[!tb] 
	\centering{
		\includegraphics[width=0.95\linewidth]{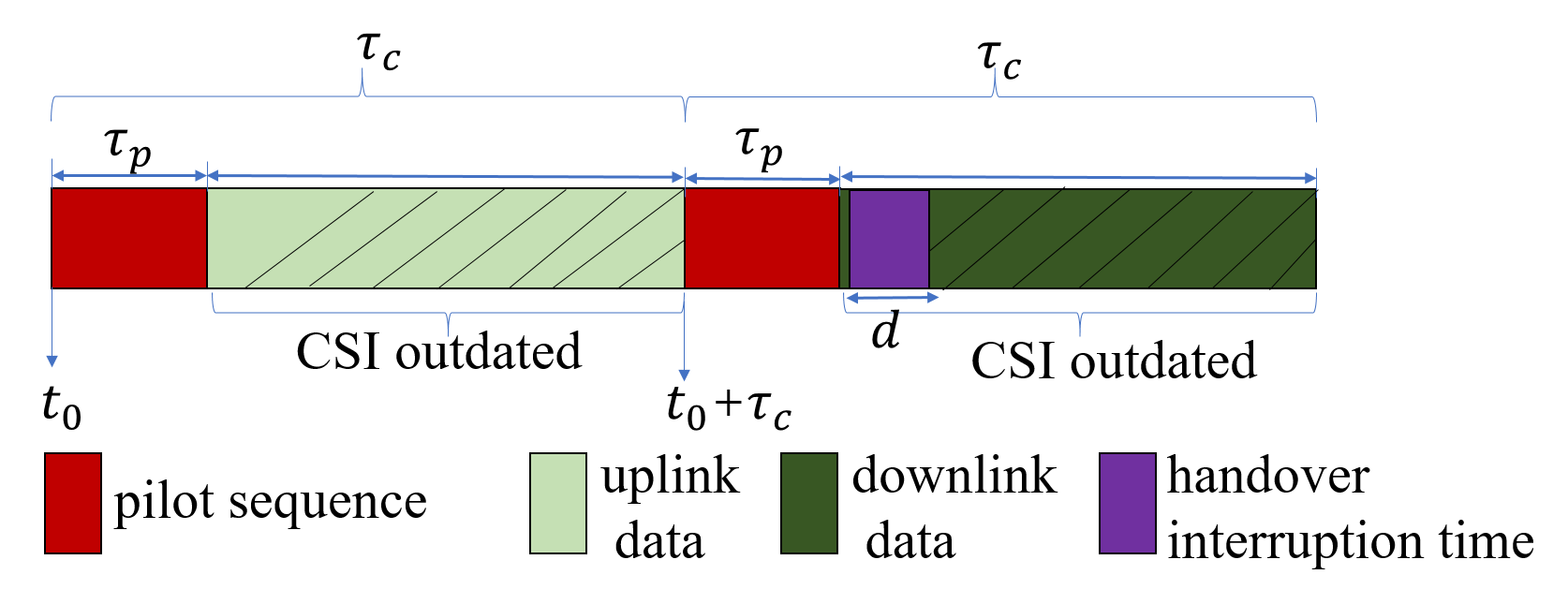}}
	\caption{Illustration of uplink and downlink communication blocks. \looseness=-1}
\label{block}
\end{figure}

We consider a time division duplex transmission as shown in Fig. \ref{block}, where the uplink and downlink data of CF-mMIMO occupy one communication block each with a total length of $\tau_c=200$ slots \cite{aging}. The channel measurement, handover decision, and signal processing only happen once per block \cite{aging}. Specifically, the network estimates the channel using a pilot sequence of $\tau_p=10$ slots within each block, while the remaining $\tau_c - \tau_p$ slots transmit data \cite{3gpp331}. We adopt random pilot assignment\footnote{Pilot assignment influences the handover algorithm performance through its impact on the quality of the AP serving set, since the channel estimates obtained from the assigned pilots determine the candidate set. However, the pilot assignment can be designed independently from the handover decision algorithm \cite{SoftHOcf} and is out of the current scope.}, where a shared pool of $\tau_p$ pilots is available network-wide and each AP randomly assigns these pilots to its served UEs \cite{MAcf}. We thus assume no pilot contamination, since random pilot assignment is known to result in low mutual interference of co-pilot UEs \cite{Rpilot}. Since the channels are only measured at the beginning of the communication block, channel information is outdated within the block and the channel realizations at different time instants in a block are different due to mobility, resulting in the channel aging effect \cite{aging}. We assume the different channel realizations are correlated and the correlation coefficient between the channel realizations at time 0 and time instant $t$ is modelled as the zeroth-order Bessel function of the first kind \cite{agingPredict}
\begin{equation}
\rho[t]=J_0\left[2 \pi \frac{v f_c}{c} T_{sa} \left(t-\tau_p-1\right)\right],
\label{rho}
\end{equation}
\noindent where $v$ is the speed of the UE, $f_c=2$~GHz is the carrier frequency, $c$ is the speed of light and $T_{sa}=0.1$~ms \cite{aging} is the channel sampling duration. The channel at time $t$ is thus \cite{aging} \looseness=-1
\begin{equation}
h_{m k}[t]=\rho[t] h_{m k}[0]+\sqrt{1-\rho^2[t]} g_{m k}[t],
\end{equation}
\noindent where $h_{m k}[0]$ is the channel at time 0, i.e., the initial state, and $g_{mk}[t] \sim N_{C}(0,R_{mk}[t])$ is the component that is independent of channel aging.

We consider both the uplink and downlink of a CF-mMIMO network. The received signal at AP $m$ for the uplink $y_m^u[t]$ and at UE $k$ for the downlink $y_k^d[t]$ at time $t$ are modelled as \looseness=-1
\begin{equation}
\begin{aligned}
& y_m^u[t]=\sum_{k=1}^K \sqrt{p_{m k}} h_{m k}[t] \epsilon_k+n_0, \\
& y_k^d[t]\!=\!\!\sum_{m=1}^M \! h_{m k}[t] \! \left(\!\sum_{i=1}^K \! \sqrt{p_{m i}} D_{mi}[t] w_{m i}[t] q_{m i} \!\! \right)\!\! +\! n_0,
\end{aligned}
\label{signalY}
\end{equation}
\noindent where $\epsilon_k \sim N_C(0,1)$ is the transmit signal from UE $k$, \mbox{$q_{mi} \sim N_C(0,1)$} is the transmit signal from AP $m$ intended for UE $i$, and $w_{mi}[t]$ is the precoding vector of AP $m$ for UE $i$ at time $t$. 

We assume minimum mean-squared error (MMSE) channel estimation $\hat{h}_{mk} \sim N_C(0, Z_{mk}[t])$ \cite{aging}, where
\begin{equation}
Z_{m k}[t]=\frac{\rho^2\left[\tau_p+1-t\right] \beta_{m k}^2[t] n_0}{p_{m k} \sum_{k \in \mathcal{P}_k} \beta_{m k}[t] n_0+p_{m k}},
\label{Qmk}
\end{equation}
\noindent with $\mathcal{P}_k$ as the UE set that uses the same pilot as UE $k$; (\ref{Qmk}) thus models the channel aging effect in channel estimation and eventually the throughput performance.

We further assume centralized partial MMSE combining and precoding, and therefore the CPU calculates the combining vector $\phi_{mk}[t]$ and precoding vector $w_{mk}[t]$ using the channel estimates $\hat{h}_{mk}$ with (18) in \cite{aging}. For uplink transmission, the serving CPU gathers the received signal $y^u_m[t]$ given by (\ref{signalY}) for all serving APs and detects the signal of UE $k$ with $\phi_{mk}[t]$. For the downlink, the CPU precodes the signal with $w_{mk}[t]$ to obtain $y^d_k[t]$.


\vspace{-0.3cm}
\subsection{Handover in CF-mMIMO}
\label{HO_CF}
\begin{figure}[!tb] 
	\centering
	\subfigure[handover procedure]{
		\label{HOproc}
		\includegraphics[width=0.8\linewidth]{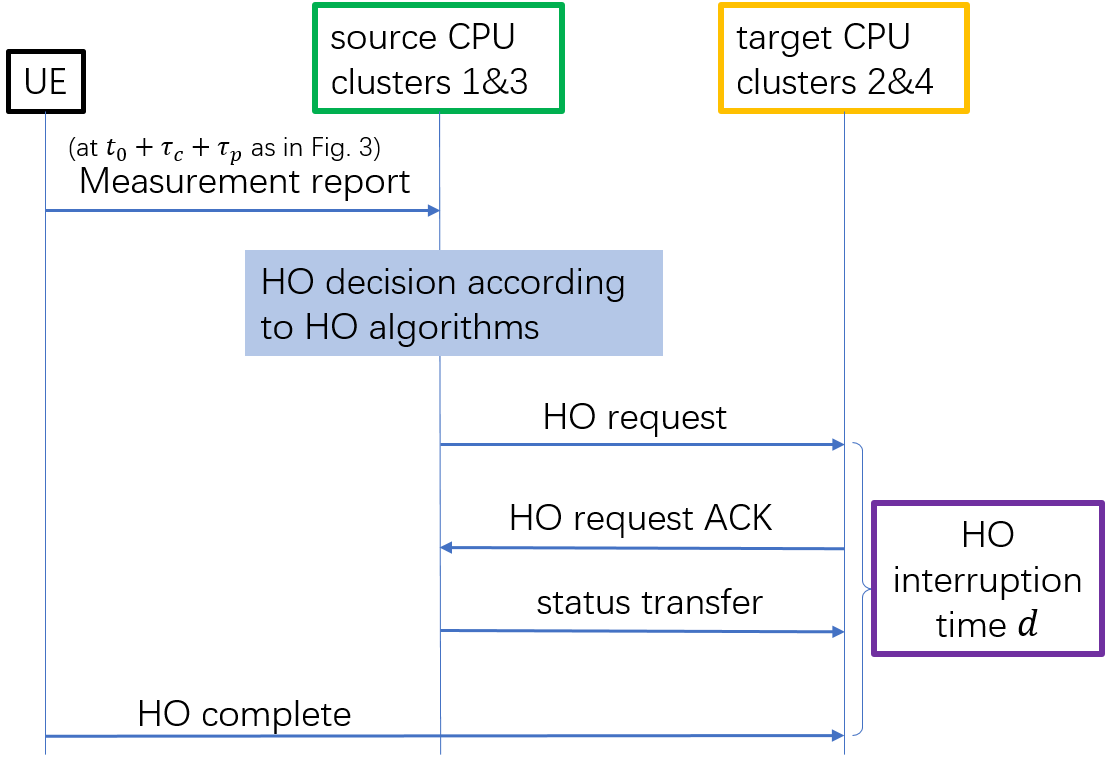}}
	\subfigure[UE mobility]{
		\label{walks}
		\includegraphics[width=0.8\linewidth]{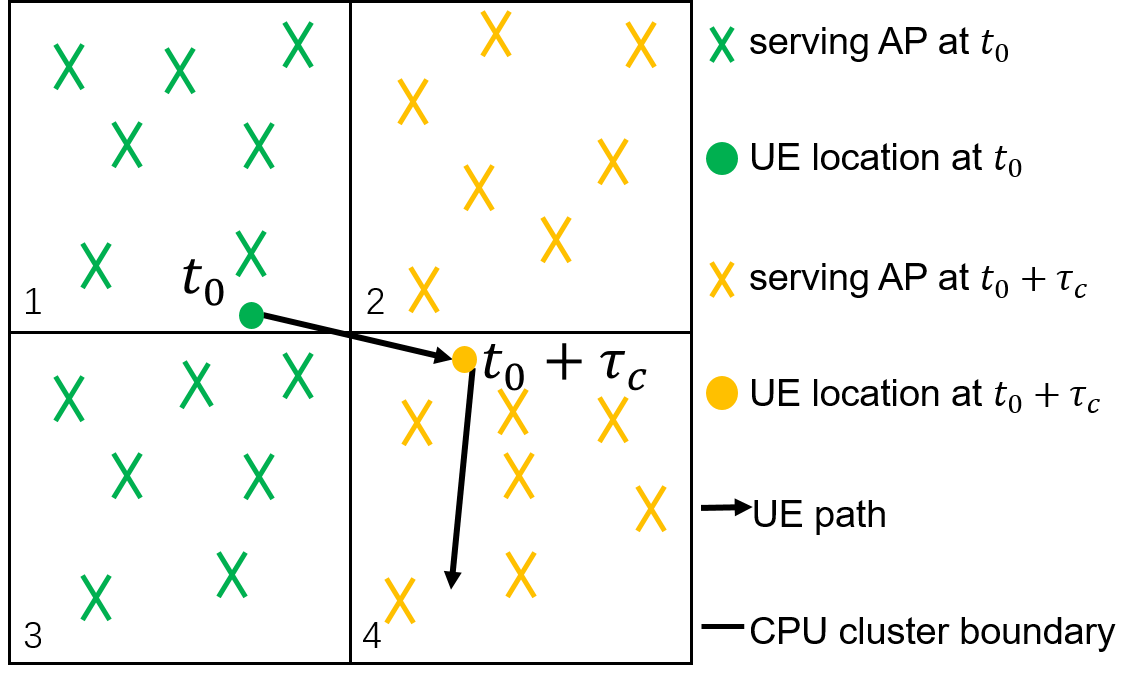}}
	\caption{Illustration of a) handover procedure and b) UE mobility for \mbox{UE-centric} CF-mMIMO, with timing as in Fig. \ref{block}.}
	\label{MobiHO}
\end{figure}

For a mobile UE in CF-mMIMO, the channel state information of the AP-UE links changes significantly over the duration of a communication block due to the channel aging effect, even though the location of the UE only slightly changes under e.g. a walking speed \cite{aging}. Therefore, the AP serving set must be determined at the beginning of each communication block, which is analogous to one synchronization signal block measurement timing configuration (SMTC) period in 5G new radio (NR) \cite{3gpp133, 3gpp331}. We consider the standard Xn/N2-based inter-base station handover in \mbox{5G-NR} \cite{3gpp300} and present its extension to \mbox{CF-mMIMO} in Fig. \ref{HOproc}. We further illustrate in Fig. \ref{walks} the mobility of an example UE in a \mbox{CF-mMIMO} network with the architecture given in Sec. \ref{Arch}. The UE starts at time instant $t=t_0$ at the position marked with the green dot and forms the initial serving AP set (green crosses) with the AP selection scheme in Sec. \ref{Arch}, i.e., the APs in CPU clusters 1 and 3. After $\tau_c$ slots, the UE has moved to the yellow dot in \mbox{Fig. \ref{walks}} and the channels are measured again at $t=t_0+\tau_c$. The candidate AP serving set (yellow crosses) denoted as $D'[t]$, i.e., the APs in CPU cluster 2 and 4, is obtained based on this new channel measurement. As per the handover procedure in Fig. \ref{HOproc}, the UE sends the measurement report of the current and candidate serving set to the source CPU clusters 1 and 3 after the channel measurement, i.e., at time $t=t_0+\tau_c+\tau_p$, and the network then makes the handover decision according to the handover algorithm. If the candidate set is selected as the new serving set, i.e., $D[t]=D'[t]$, a handover is triggered. The source serving CPU clusters then disconnect from the UE, send the handover request to the target CPU clusters 2 and 4, receive their acknowledgment, and transfer the data to the target serving CPU clusters. Finally, the handover is completed when the links between the UE and its new serving APs are established \cite{3gpp300}. During the handover procedure, the links between the UE and its new serving APs are not yet fully connected and thus cannot perform data transmission within the handover interruption time $d$, which consists of the time-to-trigger interval and handover establishment time \cite{3gpp300}. This interruption time $d$ reduces the effective throughput by shortening the usable data transmission time\footnote{We note that other make-before-break handover types, such as dual active protocol stack \cite{3gppTlat} and conditional handover \cite{3gpp300}, are designed to achieve near-zero or substantially reduced interruption times. However, these soft handover mechanisms assume a single-source-to-single-target handover paradigm and are inherently difficult to extend to distributed MIMO scenarios where a UE is simultaneously served by multiple APs. We therefore focus on the standard Xn/N2-based handover and model the handover overhead as a delay cost, deferring soft handover adaptation to future work.} within the affected uplink or downlink communication block \cite{xiao, mobilityaware, towards, tutorial}, as illustrated in Fig. \ref{block} for an example downlink block.

\subsection{Mobility-Aware Throughput Model of \mbox{CF-mMIMO}}
\label{throughput}

The impact of mobility on the throughput performance is two-fold: channel aging and handover cost \cite{myGCpaper}. We first solely consider the channel aging effect and model the throughput performance of the network without handover cost as the \textit{baseline} spectral efficiency (SE) performance $SE'$, using the performance model in \cite{aging}. We give the baseline SE of a given UE $k$ as the average SE over a communication block $n$. According to the time division duplex (TDD) transmission shown in Fig. \ref{block}, the SE performance is for the uplink when $n$ is an odd number, and for the downlink when $n$ is an even number. \looseness=-1
\begin{equation}
S E_k^{\prime}[n]\!=\!\begin{cases}\frac{1}{\tau_c}\! \sum\limits_{t=\zeta}^{t_0+n \tau_c} \log _2(1\!+\!\operatorname{SINR}_k^u[t]), n \in\{2\eta-1|\eta \in \mathbb{N}\} \\ \frac{1}{\tau_c}\! \sum\limits_{t=\zeta}^{t_0+n\tau_c} \log _2(1\!+\!\operatorname{SINR}_k^d[t]), n \in\{2\eta|\eta \in \mathbb{N}\},\end{cases}
\label{SEx}
\end{equation}
\noindent where $\zeta=t_0+(n-1) \tau_c+\tau_p+1$ is the beginning of the data transmission block, $\operatorname{SINR}_k^u[t]$ and $\operatorname{SINR}_k^d[t]$ are the \mbox{signal-to-interference-and-noise-ratio} (SINR) of UE $k$ in the the uplink and downlink, respectively, given by \cite{aging}
\begin{equation}
\operatorname{SINR}_k^u[t]\!=\!\frac{S^u}{I^u\!+\!n_0}, \;\;\; \operatorname{SINR}_k^d[t]\!=\!\frac{S^d}{I^d\!+\!n_0}.
\label{eqsinr}
\end{equation}
According to the uplink/downlink received signals in (\ref{signalY}) and the signal processing given in Sec. \ref{signal}, the desired signal strength and interference in (\ref{eqsinr}) are given by \cite{aging}
\begin{equation}
\begin{aligned}
& S^u=\rho^2\left[t-\tau_p-1\right] p_{m k} \left|\mathbb{E}\left\{ \sum_{m=1}^M \phi_{mk}^H[t] D_{m k}[t] h_{m k}[t] \right\} \right|^2 \\
& I^u=\sum_{i=1}^K p_{m i} \mathbb{E}\left\{ \left|\sum_{m=1}^M \phi_{mk}^H[t] D_{m k}[t] h_{m i}[t]\right|^2 \right\} -S^u \\
& S^d\!=\!\rho^2[t\!-\!\tau_p\!-\!1] p_{m k}\!\left| \mathbb{E}\left\{ \sum_{m=1}^M h^H_{m k}[t]  D_{m k}[t] w_{m k}[t]\right\}\right|^2 \\
& I^d=\sum_{i=1}^K p_{m i} \mathbb{E}\left\{ \left| \sum_{m=1}^M h^H_{m k}[t] D_{m i}[t] w_{m i}[t]\right|^2\right\}-S^d
\end{aligned}.
\label{eqsi}
\end{equation}
When $v=0$ and $\rho[t]=1$, $Z_{mk}[t]$ in (\ref{Qmk}) and the baseline SE in (\ref{SEx}) are the same as the static throughput model in \cite{scalableCF}.


According to the handover model in Sec. \ref{HO_CF}, we next give the comprehensive \mbox{\textit{mobility-aware}} spectral efficiency for UE $k$ at block $n$ under both channel aging and handover delay cost as \cite{xiao, mobilityaware, towards, tutorial, myGCpaper, DRLmobi}
\begin{equation}
S E_k[n]=S E_k^{\prime}[n]\left(1-x_k[n] \frac{N_k[n] d_C+\Lambda_k[n] d_{A P}}{T_{\text {block }}}\right),
\label{SEm}
\end{equation}
\noindent where $S E_k^{\prime}[n]$ is the baseline spectral efficiency given by (\ref{SEx}), $x_k[n]$ is a binary variable that states whether a handover is performed at block $n$ for UE $k$, i.e., if $x_k[n]=1$, there is a handover, otherwise $x_k[n]=0$. Furthermore, $N_k[n]$ and $\Lambda_k[n]$ are the number of CPU clusters and APs changing during the handover, respectively. For the example UE in Fig.~\ref{walks}, a handover performed at $t=t_0+\tau_c$ incurs connection changes of two CPU clusters and $2Q$ APs, i.e., $N_k[n]=2$ and $\Lambda_k[n]=2Q$. Moreover, $d_C$ is the inter-cluster handover interruption time between CPUs and $d_{AP}$ is the intra-cluster handover interruption time where there are handovers between APs. Finally, $T_\text{block}=\tau_c T_\text{slot}$ is the time for a communication block, where $T_\text{slot}$ is the time of one slot. Assuming 5G-NR numerology $\mu=0$ with a subcarrier spacing of 15 kHz, we have $T_\text{slot}=1$~ms and $T_\text{block}=$200~ms \cite{3gpp211}.

We further obtain the handover rates as the average number of handovers of UE $k$ over the mobility period of $T$ communication blocks as an important performance metric for handover algorithms, where the number of handovers is obtained by comparing the current serving AP set and the previously set at every handover decision time instant \mbox{$t=\{t_0+n\tau_c \}$}. Namely, the inter-cluster and intra-cluster handover rates for UE $k$, denoted by $H_{Ck}$ and $H_{APk}$, respectively, are given by
\begin{equation}
H_{C_k}=\frac{\sum_{n=1}^T x_k[n] N_k[n]}{T \cdot T_{\text {block}}},  H_{A P_k}=\frac{\sum_{n=1}^T x_k[n] \Lambda_k[n]}{T \cdot T_\text{block}}.
\label{HOfinal}
\end{equation}

\section{Mobility Throughput Optimization}
\label{optProblem}

We formulate the handover decision for CF-mMIMO as a novel optimization problem in Sec. \ref{optForm} and solve the transformed and relaxed problem with a near-optimal solution (\textit{nearOpt}) in \mbox{Sec. \ref{optSolve}}. We subsequently design the low complexity heuristic handover algorithm \textit{FairDiff} in Sec. \ref{myAlg}.

\subsection{Problem Formulation}
\label{optForm}
Instead of solely reducing the handover rate as in the prior literature \cite{pomdp, SoftHOcf, beerten2025}, we formulate a novel optimization problem to achieve high average \textit{nett} throughput of mobile \mbox{CF-mMIMO} that considers both channel aging and handover delay cost over the mobility period $T$. The formulation of the optimization problem is:
\begin{subequations}
\label{eq:optimizationSE}
\begin{align}
& \{x_k^*[n]\}_{n=1}^{T} = \operatornamewithlimits{arg\,max}_{\{x_k[n]\}_{n=1}^{T}} \overline{SE_k} = \frac{1}{T} \sum_{n=1}^{T} SE_k[n]
\label{objSE} \\
& \text{s.t.} \quad x_k[n] \in \{0,1\}, \quad \forall n = 1, \dots, T,
\label{constrain}
\end{align}
\end{subequations}

\subsection{Transformed \& Relaxed Solution}
\label{optSolve}

The problem \eqref{eq:optimizationSE} is an integer nonlinear program whose complexity grows exponentially with the mobility period and the network size, rendering a direct solution infeasible in our assumed large-scale CF-mMIMO network. We thus need to transform and relax it into a tractable problem. 

\textbf{Transformation 1:} according to the assumed \mbox{cluster-based} \mbox{UE-centric} (CUC) AP selection scheme in Sec.~ \ref{Arch}, illustrated in Fig. \ref{net}, the formation and change of the serving AP set are always in the unit of CPU clusters, i.e., the number of AP handovers \mbox{$\Lambda_k[n]=N_k[n]Q$}. The SE under mobility in (\ref{SEm}) can be transformed to \mbox{$SE_k[n]\!=\!SE'_k[n](1- x_k[n] N_k[n] (d_C \!+\!d_{AP} Q)/T_\text{block})$}. We note that this transformation relies on the assumption of a \mbox{cluster-based} AP selection scheme, keeping our analysis here feasible. Under alternative AP selection schemes where the UE is served by a partial APs in one CPU cluster \cite{scalableCF, DRLmaxRate, DRLmaxMinRate}, the coupling of $\Lambda_k$ and $N_k$ is scheme-dependent and complex, rendering the optimization problem in \eqref{eq:optimizationSE} intractable. Nonetheless, our heuristic handover algorithm detailed in Sec.~\ref{myAlg} is independent of the AP selection scheme; its robustness across different schemes will be shown in \mbox{Sec. \ref{algoAna:FairDiff}}. \looseness=-1


\textbf{Transformation 2:} for a UE with typical urban speed, i.e., \mbox{$v \leq 10$~m/s}, the movement during the time of a communication block $T_\text{block}=200$~ms is under two meter, which is much smaller than the length of CPU clusters (around 100~m \cite{cfsim}). Therefore, very few CPU clusters, in many cases only one, will be changed during each handover for a UE. Namely, we may assume \mbox{$N_k[n]=1, \forall n$} in the solution of (\ref{eq:optimizationSE}). Defining the constant \mbox{$Y=(d_C \!+\!d_{AP} Q)/T_\text{block}$}, the objective mobility-aware SE in \eqref{objSE} is thus transformed to $SE_k[n]=SE'_k[n] (1-Y x_k[n])$.

\textbf{Transformation 3:} the calculation of objective \eqref{objSE} requires the instantaneous SINR of each slot within a communication block. However, since the handover decision is made at the beginning of each data transmission block, the network only has access to the channel estimates obtained from the pilot segment and the SINR at the beginning of data transmission, i.e., $\operatorname{SINR}_k[\zeta]$, is the closest available value to these measurements, since channel aging has not yet degraded the estimate quality. Moreover, prior studies indicate that the SINR variation within a communication block is negligible for handover decision under typical urban mobility, even when channel aging is considered~\cite{aging, scalableaging, agingPredict, agingOptimize}. Hence, $\operatorname{SINR}_k^{u/d}[t_0+n\tau_c+\eta] \approx \operatorname{SINR}_k^{u/d}[t_0+n\tau_c]$ for all $\eta \in [0,\tau_c]$. We can therefore approximate all instantaneous SINR within a communication block with $\operatorname{SINR}_k[\zeta]$. The baseline SE in objective (\ref{objSE}) can thus be transformed to
\noindent \mbox{$SE_k'[n] \approx (\tau_c-\tau_p)/ \tau_c \log_2 (1+\operatorname{SINR}_k[\zeta])$}.

\textbf{Transformation 4:} the exact SINR given by \eqref{eqsinr} depends on the small-scale fading coefficients and the corresponding combining/precoding vectors, which require full channel state information \textit{after} handover. Therefore, it is practically infeasible to be used for handover decisions and requires transformation. Under the considered CF-mMIMO network, the number of APs is sufficiently large and the channels across different AP - UE links are assumed to be independent. Due to channel hardening, the aggregated effective channel gain becomes asymptotically deterministic. As a result, the randomness induced by \mbox{small-scale} fading averages out over independently faded AP links, allowing the SINR to be accurately approximated using only the large-scale fading coefficients \cite{MIMObook}. In this case, the desired signal power is proportional to the aggregated large-scale fading of the serving AP set, i.e., $\sum^M_{m=1} D_{mk} \beta_{mk}$, while the interference is represented by the remaining large-scale fading contributions from non-serving APs. Consequently, consistent with prior works \cite{scalableCF, aging, pomdp, SoftHOcf, beerten2025}, we transform the exact SINR in \eqref{eqsinr} to \looseness=-1
\begin{equation}
\operatorname{SINR}_k[\zeta] \approx \frac{
\sum\limits_{\mathclap{m=1}}^{\mathclap{M}} D_{mk}[\zeta]\,\beta_{mk}[\zeta]
}{
\sum\limits_{\mathclap{m=1}}^{\mathclap{M}} \,\beta_{mk}[\zeta]
- \sum\limits_{\mathclap{m=1}}^{\mathclap{M}} D_{mk}[\zeta]\,\beta_{mk}[\zeta] + 1
}.
\label{eqSimSINR}
\end{equation}
\indent Lastly, since Problem~\eqref{eq:optimizationSE} with binary variables $x_k[n] \in \{0,1\}$ involves $2^T$ possible decision sequences, we relax the constraint to the continuous interval $[0,1]$. The relaxation yields a tractable convex program whose optimal value serves as an upper bound for the original binary problem. With the transformation and relaxation above, we have:

\begin{lemma}
\label{convexSE}
The transformed and relaxed problem (\ref{eq:optimizationSE}) is convex.
\end{lemma}

\begin{proof}
At time $\zeta$, the serving AP set before handover is determined by the dynamic cooperation matrix at the previous communication block, i.e., $D_{mk}[\zeta]=D_{mk}[\zeta-\tau_c]$. Substituting this into (\ref{eqSimSINR}), we denote the transformed SINR for UE $k$ before handover as 
\begin{equation}
A = \frac{
\sum\limits_{\mathclap{m=1}}^{\mathclap{M}} D_{mk}[\zeta-\tau_c]\,\beta_{mk}[\zeta]
}{
\sum\limits_{\mathclap{m=1}}^{\mathclap{M}} \,\beta_{mk}[\zeta]
- \sum\limits_{\mathclap{m=1}}^{\mathclap{M}} D_{mk}[\zeta-\tau_c]\,\beta_{mk}[\zeta] + 1
}.
\label{eqA}
\end{equation}
\noindent Similarly, we denote the transformed SINR after handover to the candidate serving set by setting $D_{mk}[\zeta]=D'_{mk}[\zeta]$
\begin{equation}
B = \frac{
\sum\limits_{\mathclap{m=1}}^{\mathclap{M}} D'_{mk}[\zeta]\,\beta_{mk}[\zeta]
}{
\sum\limits_{\mathclap{m=1}}^{\mathclap{M}} \,\beta_{mk}[\zeta]
- \sum\limits_{\mathclap{m=1}}^{\mathclap{M}} D'_{mk}[\zeta]\,\beta_{mk}[\zeta] + 1
}.
\label{eqB}
\end{equation}
We thus transform the problem (\ref{eq:optimizationSE}) to 
\begin{subequations}
\label{eq:optimizationSEtrans}
\begin{align}
& \{x_k^*[n]\}_{n=1}^{T} = \operatornamewithlimits{arg\,max}_{\{x_k[n]\}_{n=1}^{T}} \frac{1}{T} \sum_{n=1}^{T} f(x_k[n]),
\label{objSEtrans} \\
& \text{s.t.} \quad 
 x_k[n] \in [0,1], \quad \forall n = 1, \dots, T,
\label{constraintrans}
\end{align}
\end{subequations}
where 
\begin{equation}
f(x_k)\!=\!\frac{\tau_c\!-\!\tau_p}{\tau_c} \log_2 [A+x_k(B\!-\!A)+1](1\!-\!Y x_k).
\end{equation}
The candidate serving AP set $D'[\zeta]$ at time $\zeta$ is the ideal serving set, ensuring the SINR of the candidate serving set is always higher than or equal to that of the previous set, i.e., $B \geq A$. The function $f(x_k)$ is thus a product of a positive logarithmic term $\log_2 [A+x_k(B-A)+1]$ and a negative linear term $(1-Y x_k)$, and therefore achieves the maximum value when $x_k$ satisfies
\begin{equation}
\begin{aligned}
\frac{d f\left(x_k\right)}{d x_k}= & \frac{\tau_c-\tau_p}{\tau_c} \bigg[\frac{B-A}{\left[A+x_k(B-A)+1\right] \ln 2}\left(1-Y x_k\right) \\
& -Y \log _2\left[A+x_k(B-A)+1\right] \bigg]=0
\end{aligned}.
\label{fprime}
\end{equation}
Since the objective (\ref{objSEtrans}) is the sum of $f(x_k)$, the transformed problem is convex.
\end{proof}


This lemma gives the solution as a transcendental equation. We solve (\ref{fprime}) using Newton's method \cite{Newton}:
\[
x_k^{(n+1)} = x_k^{(n)} - \frac{f'\big(x_k^{(n)}\big)}{f''\big(x_k^{(n)}\big)}.
\]
The iteration is initialized with \(x_k^{(0)}=0.5\) and terminated when\[
|x_k^{(n+1)}-x_k^{(n)}|<\varepsilon\quad\text{or}\quad |f'(x_k^{(n+1)})|<\varepsilon,
\] where \(\varepsilon=10^{-6}\) \cite{NewtonComplex} is a prescribed tolerance. We denote the solution as $x_k=C \in \mathbb{R}$ and convert it back to an integer satisfying the constraint (\ref{constrain}). We give the conversion policy as follows: if $C<0$, $f(x_k)$ monotonically decreases on $[0,1]$ and the objective (\ref{objSE}) is achieved with $x_k=0$; if $C>1$, $f(x_k)$ monotonically increases on $[0,1]$ and the objective (\ref{objSE}) is achieved with $x_k=1$; if $0<C<1$, $x_k=C$ is rounded to the closest integer. 

By thus solving the transformed and relaxed optimization problem (\ref{eq:optimizationSEtrans}), we derive the near-optimal handover algorithm \textit{nearOpt} presented in Alg. \ref{optimal}, which can be used to evaluate the performance of our \textit{FairDiff} algorithm. We emphasize that the transformation above serves solely to make the optimization problem (\ref{eq:optimizationSE}) tractable. We will use the exact signal and throughput model given in \mbox{Sec. \ref{model}} in all performance evaluations in \mbox{Sec. \ref{results}}. Moreover, we will in Sec. \ref{algoAna:nearOpt} validate Transformations 2-4 and compare our \textit{nearOpt} algorithm against a \mbox{genie-aided} scheme that evaluates the exact per-block throughput in~\eqref{SEm} and selects the binary decision that yields the higher value to validate its near-optimality.

\begin{algorithm}[t]
\caption{Near-Optimal Handover Algorithm (\textit{nearOpt})} \label{optimal}
\textbf{Input:} The previous SNR $\beta_{mk} [\zeta-\tau_c]$ and the current SNR $\beta_{mk} [\zeta]$ between AP $m$ and UE $k$ for \mbox{$m=1,...,M$} and $k = 1,...,K$. The previous dynamic cooperation matrix $D [\zeta-\tau_c]$ and the candidate dynamic cooperation matrix $D' [\zeta]$. The cluster handover interruption time $d_C$. 
\begin{algorithmic}[1]

\For {$k = 1,...,K$}

\State Solve the transcendental equation (\ref{fprime}) with Newton's method and obtain the solution $x_k=C$.

\If {$C < 0$ \textbf{or} $\mathrm{round}(C) = 0$}

\State Set \mbox{$D_{mk} [\zeta]\!\! =\!\! D_{mk} [\zeta\!-\!\tau_c]$},for $m\!\!=\!\!1,...,M$ (no HO).\looseness=-1

\ElsIf {$C > 1$ \textbf{or} $\mathrm{round}(C) = 1$}

\State Set $D_{mk} [\zeta] = D'_{mk} [\zeta]$, for \mbox{$m = 1,...,M$} (HO).

\EndIf

\EndFor

\end{algorithmic}
\textbf{Output:} The dynamic cooperation matrix $D$.
\end{algorithm}


\section{Fairness-Differentiated Handover Scheme}
\label{algo}
We propose in Sec. \ref{myAlg} a fairness-differentiated handover algorithm,  \textit{FairDiff}, as a heuristic solution with low computational complexity for the optimization problem in \mbox{Sec. \ref{optForm}}. We also present the reference benchmark handover schemes in Sec. \ref{baseline}. 

\subsection{FairDiff: Handover Algorithm Based on Fairness Index}
\label{myAlg}

The \textit{nearOpt} algorithm in Sec. \ref{optSolve} requires solving a transcendental equation, which entails high computational complexity. In order to reduce the complexity and propose a practical solution for mobile CF-mMIMO, we propose in \mbox{Alg. \ref{network}} a novel fairness-differentiated handover scheme \textit{FairDiff} for mobile \mbox{CF-mMIMO} as the heuristic solution to the optimization problem in (\ref{eq:optimizationSE}). We design \textit{FairDiff} by considering the performance trade-off described by (\ref{SEm}). To achieve high performance of \mbox{CF-mMIMO} under mobility, i.e., high $SE_k$, the handover algorithm should conduct necessary handovers to maintain high baseline throughput under channel aging, i.e., high $SE'_k$. On the other hand, it should also reduce the number of handovers to decrease handover delay cost by reducing unnecessary handovers, i.e., $x_k=0$ for more communication blocks. To this end, our algorithm differentiates between necessary handovers to be prioritized and optional handovers to be minimized. For UEs requiring urgent handovers, we apply a liberal handover policy to encourage necessary handovers. A strict handover policy is applied to UEs that do not require urgent handovers to reduce the handover interruption time cost. We do so by applying different handover policies for the UEs above and below a \textit{\mbox{handover-policy-differentiation} threshold} $\alpha$ obtained via Jain's fairness index, and thus name our handover algorithm \mbox{fairness-differentiated} (\textit{FairDiff}).


\begin{algorithm}[t]
\caption{Proposed Fairness-Differentiated Handover Algorithm (\textit{FairDiff})} \label{network}
\textbf{Input:} The previous SNR $\beta_{mk} [t-\tau_c]$ and the current SNR $\beta_{mk} [t]$ between AP $m$ and UE $k$ for \mbox{$m=1,...,M$} and $k = 1,...,K$. The previous dynamic cooperation matrix $D [t-\tau_c]$ and the candidate dynamic cooperation matrix $D' [t]$.
\begin{algorithmic}[1]

\State Calculate the Jain's fairness index $F$ with (\ref{fi}).

\State Sort the current SNR $s^{cur}_k$ for all UEs in ascending order.

\State Set the handover policy threshold $\alpha = s^{cur}_{\left\lceil ((1-F)K)^{th} \right \rceil}$.

\For {$k = 1,...,K$}

\State Calculate the previous total SNR $s^{bef}_k$ for UE $k$, $s^{bef}_k = \sum{\beta_{mk} [t-\tau_c] } D_{mk} [t-\tau_c]$, for $m=1,...,M$.

\State Calculate the current total SNR without the change of serving set $s^{cur}_k$ for UE $k$, $s^{cur}_k = \sum{\beta_{mk} [t] } D_{mk} [t-\tau_c]$, for $m=1,...,M$.

\State Calculate the total SNR with the candidate serving set $s^{new}_k$ for UE $k$, $s^{new}_k = \sum{\beta_{mk} [t] } D'_{mk} [t]$, for \mbox{$m=1,...,M$}.

\If {$s^{cur}_k < \alpha$}

\If {$s^{new}_k > s^{cur}_k + \gamma_1 $} 

\State Set $D_{mk} [t] = D'_{mk} [t]$, for $m = 1,...,M$ (HO).

\Else   \;\;\;\;\;\;\;\;\;\;\% no HO

\State Set \mbox{$D_{mk} [t]\!\! =\!\! D_{mk} [t\!-\!\tau_c]$}, for $m\!\!=\!\!1,...,M$.

\EndIf

\Else

\If {$s^{new}_k > s^{cur}_k + \gamma_1 $ \& $s^{cur}_k < s^{bef}_k - \gamma_2 $}
\State Set $D_{mk} [t] = D'_{mk} [t]$, for $m = 1,...,M$ (HO).
\Else     \;\;\;\;\;\;\;\;\;\;\% no HO
\State Set \mbox{$D_{mk} [t]\!\! =\!\! D_{mk} [t\!-\!\tau_c]$}, for $m\!\!=\!\!1,...,M$.
\EndIf

\EndIf

\EndFor

\end{algorithmic}
\textbf{Output:} The dynamic cooperation matrix $D$.
\end{algorithm}

In Alg. \ref{network}, the UE-specific handover policy is chosen by comparing the UE's total SNR of all serving APs against the threshold that reflects the SNR performance of the whole network, as calculated in Steps 1 - 3. We design $\alpha$ aiming to preserve the premise of uniform performance for all UEs in \mbox{CF-mMIMO} \cite{cfvs}, by enabling good channel conditions from the serving AP set for all UEs. Therefore, we use Jain's fairness index $F$, which is a well-established metric \cite{shortFI, optimalFI} used to quantify the uniformity of resource allocation, to measure the fairness in terms of the serving AP set performance $s_k$, \looseness=-1
\begin{equation}
F=\frac{\left(\sum^{K}_{k=1} s_k\right)^2}{K \cdot \sum^K_{k=1} s_k^2},
\label{jain}
\end{equation}
\noindent The exact performance of the serving set is represented by the exact SINR given by \eqref{eqsinr}. However, computing the exact SINR at the handover decision moment requires knowledge of the full channel states and handover decisions of all UEs across the network, rendering prohibitive signaling overhead and computational complexity, especially in a large-scale CF-mMIMO network. The average SNR that reflects the received signal strength from the serving AP set is thus widely used as a low-complexity proxy for service quality in existing CF-mMIMO studies \cite{scalableCF, cfsim, aging}. Consequently, consistent with prior works \cite{pomdp, SoftHOcf, beerten2025}, we use the total SNR for UE $k$ of all serving \mbox{AP - UE} links to represent the serving AP set quality at time $t$, i.e., \mbox{$s_k = \sum_{m=1}^M \beta_{m k}[t] D_{m k}[t-\tau_c]$}, to design a low-complexity handover algorithm for practical CF-mMIMO networks. Substituting the SNR into (\ref{jain}), we obtain the fairness index for SNR as
\begin{equation}
F=\frac{ (\sum_{k=1}^K \sum_{m=1}^M \beta_{m k}[t] D_{m k}[t-\tau_c])^2}{K \sum_{k=1}^K (\sum_{m=1}^M \beta_{m k}[t] D_{m k}[t-\tau_c])^2}.
\label{fi}
\end{equation}
\noindent The values of $F$ lie in $[1/K, 1]$; when $F \to 1$, the UEs achieve similar throughput over the network and the system is considered to be fair, aligning with the promise of uniform throughput in CF-mMIMO. In Steps 2 - 3, threshold $\alpha$ is calculated based on $F$. Next, in Steps 4 - 7, the UE's previous total SNR $s^{bef}_k$, the UE's current total SNR $s^{cur}_k$ with the connected serving set, and the UE's total SNR with the candidate serving set $s^{new}_k$ are calculated for further comparison. The following steps differentiate the handover policy according to the current SNR of the UE. Since $\alpha$ corresponds to the $\left\lceil ((1-F)K)^{th} \right \rceil$ smallest SNR among all UEs, it represents the boundary below which the \mbox{worst-served} $(1-F)K$ UEs fall, i.e., the fraction that lacks fairness. These UEs are thus considered to be not fairly served and would benefit from handovers to switch to better serving sets for a better SNR. Consequently, a UE with a current SNR worse than the threshold $\alpha$ is considered as poorly-served and requires a handover urgently. A liberal handover policy is applied in Steps 8 - 14 of Alg. \ref{network}, where a handover will be triggered as long as the candidate serving set can provide a better SNR (over a margin $\gamma_1$). The other UEs whose SNR is above $\alpha$ are considered well-served and the handover requirement is not urgent. Therefore, a stricter handover policy than in Steps \mbox{8 - 14} is applied in Steps 15 - 22. We apply the classic hysteresis policy \cite{HOns3} for the \mbox{well-served} UEs. Besides the condition that the candidate serving set should provide better SNR, the UE's current total SNR should also drop below the previous SNR (over a margin $\gamma_2$), in order to trigger a handover. \looseness=-1

\subsection{Reference Handover Algorithms}
\label{baseline}

\begin{algorithm}[tb]
\caption{POMDP AP Selection and Handover Algorithm (\textit{POMDP}) \cite{pomdp}} \label{pomdp}
\textbf{Input:} The previous SNR $\beta_{mk} [t-\tau_c]$ and the current SNR $\beta_{mk} [t]$ between AP $m$ and UE $k$ for \mbox{$m=1,...,M$} and $k = 1,...,K$. The previous dynamic cooperation matrix $D [t-\tau_c]$.

\begin{algorithmic}[1]

\State Achieve the candidate serving set $D'[t]$ using the POMDP framework in \cite{pomdp} by choosing an action of UE-AP connection that maximize the reward function in (27) in \cite{pomdp} , i.e., perform Steps 3 to 8 in Algorithm 3 in \cite{pomdp}.

\State Achieve the SNR-based classification threshold $\beta_{\text{th}}$ with (29) in \cite{pomdp}.

\For {$k = 1,...,K$}

\If {$s^{cur}_k \leq \beta_{\text{th}}$}
\State Set $D_{mk} [t] = D'_{mk} [t]$, for \mbox{$m = 1,...,M$} (HO).
\Else
\State Set \mbox{$D_{mk} [t]\!\! =\!\! D_{mk} [t\!-\!\tau_c]$}, for $m\!\!=\!\!1,...,M$ (no HO).
\EndIf

\EndFor

\end{algorithmic}
\textbf{Output:} The dynamic cooperation matrix $D [t]$ at time $t$.
\end{algorithm}

\begin{algorithm}[tb]
\caption{Hysteresis Handover Algorithm (\textit{hysteresis}) \cite{beerten2025}} \label{hysteresis}
\textbf{Input:} The previous SNR $\beta_{mk} [t-\tau_c]$ and the current SNR $\beta_{mk} [t]$ between AP $m$ and UE $k$ for \mbox{$m=1,...,M$} and $k = 1,...,K$. The previous dynamic cooperation matrix $D [t-\tau_c]$ and the candidate dynamic cooperation matrix $D' [t]$.
\begin{algorithmic}[1]

\For {$k = 1,...,K$}

\State Perform Steps 5 to 7 in Algorithm \ref{network}

\If {$s^{new}_k > s^{bef}_k + \delta_1 $ and $s^{cur}_k < s^{bef}_k - \delta_2 $}
\State Set $D_{mk} [t] = D'_{mk} [t]$, for \mbox{$m = 1,...,M$} (HO).
\Else
\State Set \mbox{$D_{mk} [t]\!\! =\!\! D_{mk} [t\!-\!\tau_c]$}, for $m\!\!=\!\!1,...,M$ (no HO).
\EndIf

\EndFor

\end{algorithmic}
\textbf{Output:} The dynamic cooperation matrix $D [t]$ at time $t$.
\end{algorithm}


\begin{algorithm}[t]
\caption{UE-Performance-Aware Handover Algorithm (\textit{UPA}) \cite{SoftHOcf}} \label{drop}
\textbf{Input:} The previous SNR $\beta_{mk} [t-\tau_c]$ and the current SNR $\beta_{mk} [t]$ between AP $m$ and UE $k$ for \mbox{$m=1,...,M$} and $k = 1,...,K$. The previous dynamic cooperation matrix $D [t-\tau_c]$ and the candidate dynamic cooperation matrix $D' [t]$.
\begin{algorithmic}[1]

\For {$k = 1,...,K$}

\State Perform Steps 5 and 6 in Algorithm \ref{network}

\If {$s^{cur}_k < s^{bef}_k - \theta$}
\State Set $D_{mk} [t] = D'_{mk} [t]$, for \mbox{$m = 1,...,M$} (HO).
\Else 
\State Set \mbox{$D_{mk} [t]\!\! =\!\! D_{mk} [t\!-\!\tau_c]$}, for $m\!\!=\!\!1,...,M$ (no HO).
\EndIf

\EndFor

\end{algorithmic}
\textbf{Output:} The dynamic cooperation matrix $D [t]$ at time $t$.
\end{algorithm}



We briefly present the reference benchmark handover algorithms from \cite{pomdp, beerten2025, SoftHOcf} in this section. The \textit{POMDP} algorithm proposed in \cite{pomdp} is described in Alg. \ref{pomdp}. It first determines the candidate serving set after handover using a POMDP framework given in (17) of \cite{pomdp} in Step 1. The state space of this POMDP framework is formed by partially observing the channel state of the serving APs to a given UE and the action space is all possible serving APs after handover. The action that maximizes a baseline SE-based reward function given by (27) in \cite{pomdp} is then chosen to be the candidate serving set. Compared to the serving AP set selection in \mbox{Sec. \ref{Arch}}, the POMDP framework forms the candidate set with less channel state information, but much higher complexity \cite{pomdp}. Alg. \ref{pomdp} additionally reduces the number of handovers by setting a \mbox{SNR-based} threshold $\beta_{\text{th}}$ in Step 2, then \mbox{Steps 3 - 9} decide whether to keep the previous serving set or to change it according to the threshold. The \textit{hysteresis} handover algorithm proposed in \cite{beerten2025} is described in Alg. \ref{hysteresis}. This algorithm monitors the SNR of the individual UE at a given time instant in Step 2 and introduces the hysteresis margins in Steps \mbox{3 - 7}. Therefore, to trigger a handover, the UE's current SNR must have deteriorated from the previous SNR over a margin $\delta_1$ and the new SNR obtained by the candidate serving set should also be better than the previous SNR over a margin $\delta_2$. The \textit{UE-performance-aware (UPA)} handover scheme in \cite{SoftHOcf} is described in Alg. \ref{drop}. The algorithm only compares the UE's previous SNR with the current SNR and triggers a handover when the difference exceeds a margin $\theta$.

Besides the handover algorithm benchmarks, we also consider two boundary cases --  \textit{original CF-mMIMO} and \mbox{\textit{always-handover}} -- as references. The \textit{original CF-mMIMO} case serves all UEs with all APs in the network, and thus incurs only channel aging but no handover at all during the entire UE mobility path, representing the theoretical upper bound for the throughput of a mobile \mbox{CF-mMIMO} network. The \textit{always-handover} case is where the UE is always served by the ideal candidate serving set according to a given AP selection scheme. This case presents the performance upper bound for the baseline SE in \eqref{SEx} without handover cost of scalable mobile CF-mMIMO for a given serving set size and AP selection scheme, but also results in the highest handover rate. \looseness=-1

\section{Results}
\label{results}

In Sec. \ref{results:ref} we first analyse the throughput performance of prior benchmark handover schemes given in Sec. \ref{baseline} in mobile CF-mMIMO. We then in Sec. \ref{results:myAlgo} evaluate the performance of our proposed \mbox{\textit{nearOpt}} and \textit{FairDiff} handover algorithms for \mbox{CF-mMIMO} proposed in Sec.~\ref{optSolve} and Sec.~\ref{myAlg}, respectively, under both PPP and realistic urban mobile CF-mMIMO networks. We finally in Sec. \ref{algoAna} present a detailed analysis of our proposed algorithms.\looseness=-1

We obtain the network distribution and UE mobility patterns according to the models in Sec. \ref{Arch}. We study the UE-centric CF-mMIMO architecture with the CUC serving AP selection scheme stated in Sec. \ref{Arch}, with $\{Q,E\} = \{20,7\}$ to obtain an average serving AP set size $G=34$. We simulate all $K=50$ UE tracks over a mobility period of $(T \cdot T_\text{block})=1000$~s for the pedestrian UEs with $v=1$ m/s and $(T \cdot T_\text{block})=100$~s for the vehicular UEs with $v=10$ m/s. Since $T_\text{block}=200$~ms (\textit{cf. }Sec.~\ref{throughput}), the entire mobility period consists of $T=5000$ and \mbox{$T=500$} communication blocks for $v=1$ m/s and $v=10$ m/s, respectively. We thereby obtain a distribution of UE throughput over $50 \times (5000/0.2)=1250000$ and $50 \times (500/0.2)=125000$ Monte-Carlo realizations for pedestrian and vehicular UEs, respectively. We consider the handover interruption times $d_C=0.1$~s and $d_{AP}=0.02$~s to be constant values within the typical range (10 - 100~ms) specified for 5G-NR handover interruption~\cite{differdelay, mobilityaware, DRLmobi, 3gpp133}. For the reference benchmarks, we consider $\delta_1=\delta_2=4$~dB and $\theta=1$~dB, the \mbox{best-tuned} parameters as given in \cite{pomdp, beerten2025, SoftHOcf}. For our proposed \textit{FairDiff} algorithm, we consider the \mbox{best-tuned} margin \mbox{$\gamma_1=\gamma_2=1$ dB} and detail the margin tuning in Sec. \ref{algoAna:FairDiff}.


\subsection{Performance of Mobile CF-mMIMO}
\label{results:ref}

\begin{figure}[!tb] 
	\centering
	\includegraphics[width=1\columnwidth]{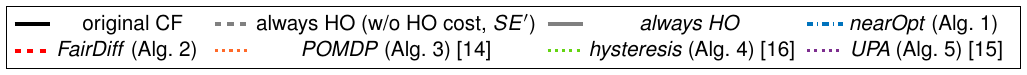}
	\subfigure[mobility-aware SE per UE]{
	\label{PPPuec.mobiSE}
	\includegraphics[width=1\columnwidth]{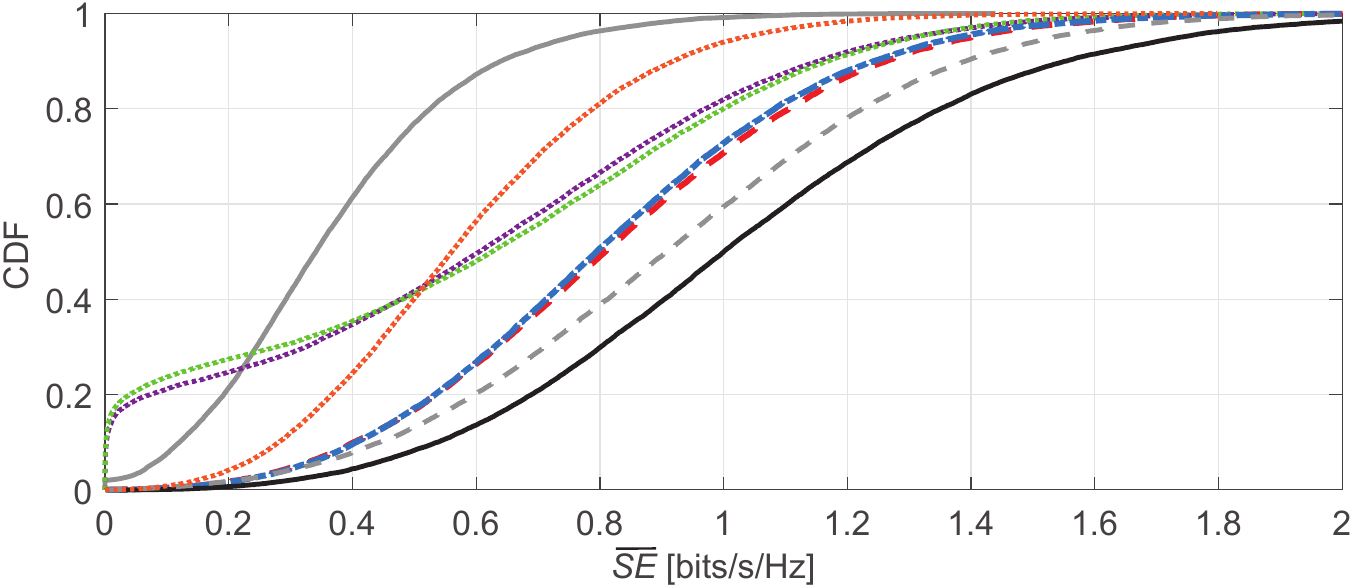}}
	\subfigure[baseline SE per UE]{
	\label{PPPuec.baseSE}
		\includegraphics[width=0.48\linewidth]{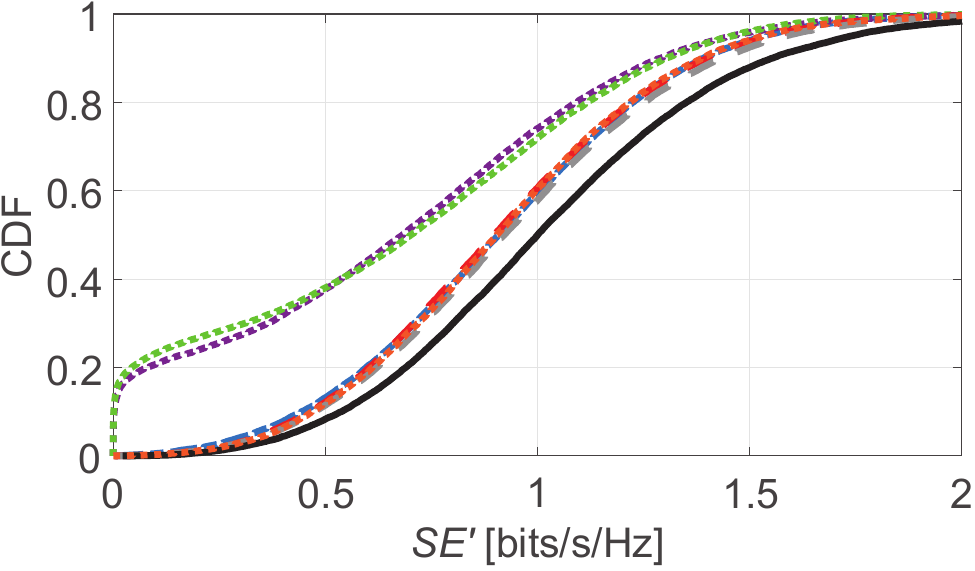}}
	\subfigure[inter-cluster handover rate per UE]{
	\label{PPPuec.HO}
		\includegraphics[width=0.475\linewidth]{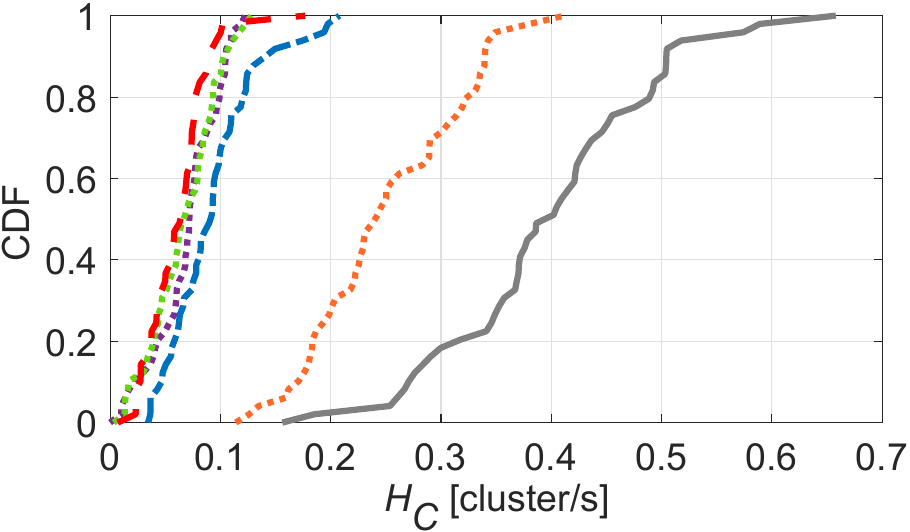}}
	\caption{Throughput performance and handover rate of \mbox{CF-mMIMO} with different handover schemes (PPP network, RWP UE mobility, $v = 1$ m/s).}
	\label{PPPuec}
\end{figure}

\begin{figure}[!t] 
	\centering
	\subfigure[$SE'$, \textit{always-HO} \& \textit{no-HO}]{
		\label{SEvT.sub.1}
		\includegraphics[width=0.48\linewidth]{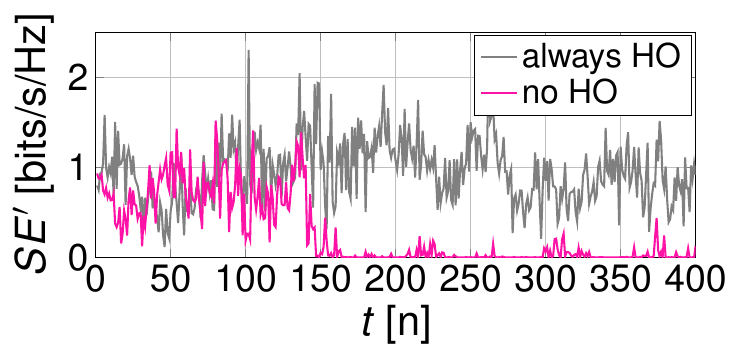}}
	\subfigure[$N_k$, \textit{always-HO}]{
		\label{SEvT.sub.5}
		\includegraphics[width=0.48\linewidth]{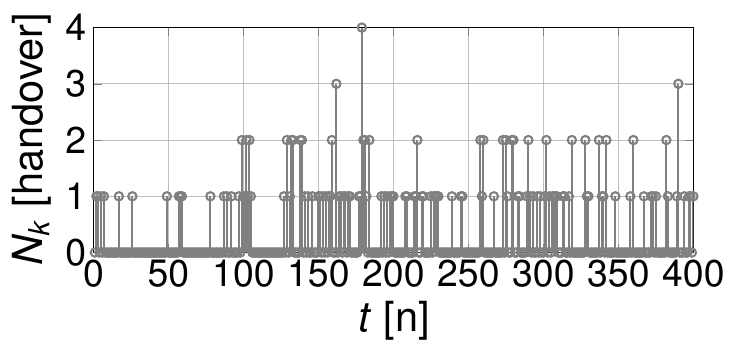}}
    \subfigure[$SE'$, \textit{POMDP} (Alg. \ref{pomdp})]{
		\label{SEvT.sub.pomdp}
		\includegraphics[width=0.48\linewidth]{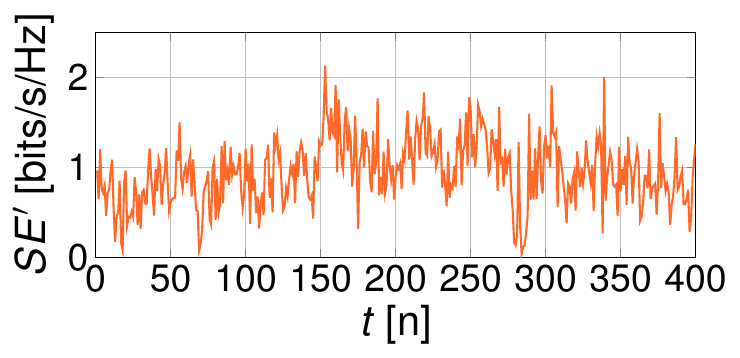}}
	\subfigure[$N_k$, \textit{POMDP} (Alg. \ref{pomdp})]{
		\label{SEvT.sub.pomdpHO}
		\includegraphics[width=0.48\linewidth]{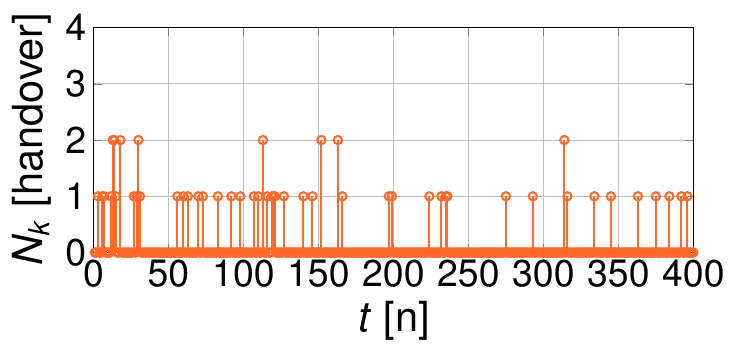}}
    \subfigure[$SE'$, \textit{hysteresis} (Alg. \ref{hysteresis}) (\textit{UPA} gives similar results)]{
		\label{SEvT.sub.2}
		\includegraphics[width=0.48\linewidth]{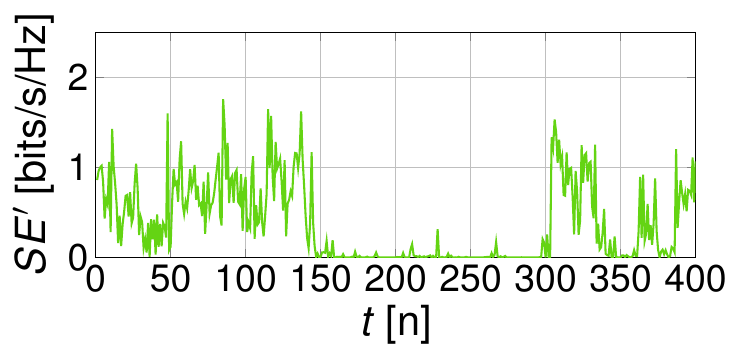}}
	\subfigure[$N_k$, \textit{hysteresis} (Alg. \ref{hysteresis}) (\textit{UPA} gives similar results)]{
		\label{SEvT.sub.6}
		\includegraphics[width=0.48\linewidth]{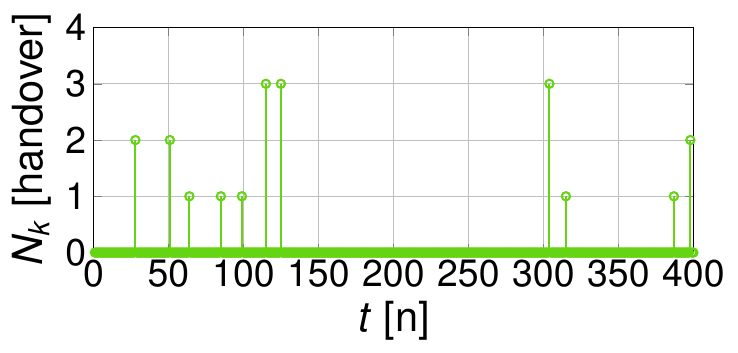}}
	\caption{Baseline SE $SE'$ and number of handovers $N_k$ of an example UE vs. time in CF-mMIMO with benchmark handover schemes (PPP network, RWP UE mobility, \mbox{$v = 1$ m/s}).}
	\label{SEvT}
\end{figure}

In this section, we study the throughput performance of \mbox{CF-mMIMO} under our comprehensive mobility-aware throughput model given in \eqref{SEm}, in order to establish the necessity of our problem formulation in Sec.~\ref{optProblem} and our proposed \textit{nearOpt} and \textit{FairDiff} handover algorithms. We consider the PPP distributed network with the pedestrian UE speed \mbox{$v=1$ m/s} and will extend our studies to other networks in Sec. \ref{otherReal}.  Fig. \ref{PPPuec} presents the throughput performance and handover rate of different handover schemes and boundary cases for \mbox{CF-mMIMO}. Specifically, \mbox{Fig. \ref{PPPuec.mobiSE}} shows the spectral efficiency $SE$ given by our throughput model in (\ref{SEm}), which considers both the channel aging effect and handover interruption time cost (plus for reference, the baseline $SE'$ without handover cost for \mbox{\textit{always-handover}}). In order to examine the impact on the overall throughput performance of these two components, \mbox{Fig. \ref{PPPuec.baseSE}} shows the baseline spectral efficiency $SE'$ only considering channel aging effect as given in (\ref{SEx}), whereas \mbox{Fig. \ref{PPPuec.HO}} shows the \mbox{inter-cluster} handover rate $H_C$ as the main contributor to the handover cost since $d_C \gg d_{AP}$. We furthermore present in \mbox{Fig. \ref{SEvT}} the baseline SE and number of handovers of an example UE versus time over its mobility path to directly illustrate how the handovers are triggered with different handover schemes. We use the SE of the \mbox{\textit{no-handovers}} case, where no handover is performed during this path, as a reference to illustrate the moment that requires handovers. \looseness=-1

Let us start by establishing the basic behavior of \mbox{UE-centric} \mbox{CF-mMIMO} under mobility. Firstly, Fig. \ref{PPPuec.mobiSE} shows that if the handover interruption time is not considered (i.e. baseline $SE'$), the \textit{always-handover} case achieves similarly high throughput compared to the \textit{original CF-mMIMO} case, since the ideal serving AP set always provides the best channel conditions. However, Fig. \ref{PPPuec.mobiSE} shows that when handover cost is taken into account, the nett mobile SE of scalable \mbox{CF-mMIMO} with \textit{always-handover} is significantly degraded compared to the original CF upper bound. Namely, the \mbox{\textit{always-handover}} case with handover cost obtains significantly worse SE performance than the case only with channel aging, due to its high handover rate as shown in Fig. \ref{PPPuec.HO}. \mbox{Figs. \ref{SEvT.sub.1}} and \ref{SEvT.sub.5} further show that this high handover rate comes from many unnecessary handovers. For the example UE considered in Fig. \ref{SEvT}, \mbox{Fig. \ref{SEvT.sub.1}} shows that the performance of the \mbox{\textit{no-handovers}} case (i.e., keep the initial serving AP set over the entire mobility period) only degrades a little compared to the \mbox{\textit{always-handover}} case over the period of $t=\{0-150n\}$, which means it is unnecessary to conduct so many handovers during this period. However, \mbox{Fig. \ref{SEvT.sub.5}} shows that \mbox{\textit{always-handover}} performs frequent handovers over the whole period, regardless of their necessity as revealed by Fig. \ref{SEvT.sub.1}. Namely, triggering a handover every time to always maintain the ideal candidate serving set leads to unnecessarily high handover cost, emphasizing the vital importance of a handover scheme to select necessary handovers to improve the performance of mobile UE-centric CF-mMIMO.

Let us next study the performance of the prior benchmark handover schemes from Sec. \ref{baseline}. \mbox{Fig. \ref{PPPuec.baseSE}} shows that the \textit{POMDP} scheme achieves similar baseline SE performance as the \mbox{\textit{always-handover}} upper bound, because the reward function of the POMDP framework is designed to maximize the baseline SE \cite{pomdp}. However, due to its focus on baseline SE optimization, \mbox{Fig. \ref{SEvT.sub.pomdpHO}} shows that although it performs the necessary handover at $t=150n$, it still fails to suppress the unnecessary handovers for other moments and thus still incurs a significantly higher handover rate than the other selective handover schemes (\textit{cf.} Fig. \ref{PPPuec.HO}). As a result, \mbox{Fig. \ref{PPPuec.mobiSE}} shows that \textit{POMDP} obtains only half of the median throughput of the baseline SE of \mbox{\textit{always-handover}} under mobility. Fig. \ref{PPPuec.HO} shows that the \mbox{\textit{UPA}} and \textit{hysteresis} schemes that solely reduce the handovers achieve significantly lower handover rate than \textit{POMDP} and \mbox{\textit{always-handover}}. However, they both achieve significantly low throughput performance in terms of \textit{both} SE under mobility (\mbox{\textit{cf.} Fig. \ref{PPPuec.mobiSE}}) and baseline SE (\textit{cf. }\mbox{Fig. \ref{PPPuec.baseSE}}). Moreover, \mbox{Fig. \ref{PPPuec.mobiSE}} shows that using these prior handover schemes, about 15\% of the worst-served UEs remain in outage and about 30\% and 40\% of the \mbox{worst-served} UEs achieve significantly worse SE performance than the \mbox{\textit{always-handover}} case and \textit{POMDP}, respectively. Crucially, Fig. \ref{SEvT.sub.6} shows that this is because the \textit{hysteresis} and \textit{UPA} schemes fail to trigger necessary handovers at the moment needed due to the indiscriminate handover policy across all UEs, leading to a large throughput degradation at $t=150n$ shown in \mbox{Fig. \ref{SEvT.sub.2}}. As a result, the worst-served UEs that require handovers urgently cannot get the chance to improve their performance with a serving set with better channel conditions.

In summary, our results in Figs. \ref{PPPuec} and \ref{SEvT} reveal that none of the prior benchmarks achieve uniformly good throughput performance for mobile CF-mMIMO. Their poor performance is brought by their fundamentally biased objective, i.e., to solely optimize the baseline SE considering only channel aging (as in \textit{POMDP}) or to solely reduce the handover rate (as in \textit{hysteresis} and \textit{UPA}). This emphasizes the need for a new handover algorithm that optimizes the nett throughput given by the comprehensive mobility-aware throughput model in \eqref{SEm}, in order to restore the uniformly good performance of \mbox{CF-mMIMO} under mobility.

\subsection{Performance of Proposed Handover Algorithms}
\label{results:myAlgo}

\subsubsection{In an Example PPP Network}
\label{myAlgoPPP}

\begin{figure}[!t] 
	\centering
	\subfigure[$SE'$, \textit{FairDiff} (Alg. \ref{network})]{
		\label{SEvTb.sub.fair}
		\includegraphics[width=0.48\linewidth]{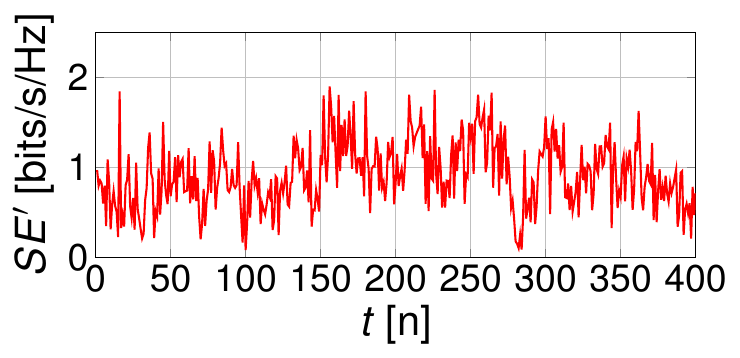}}
	\subfigure[$N_k$, \textit{FairDiff} (Alg. \ref{network})]{
		\label{SEvTb.sub.fairHO}
		\includegraphics[width=0.48\linewidth]{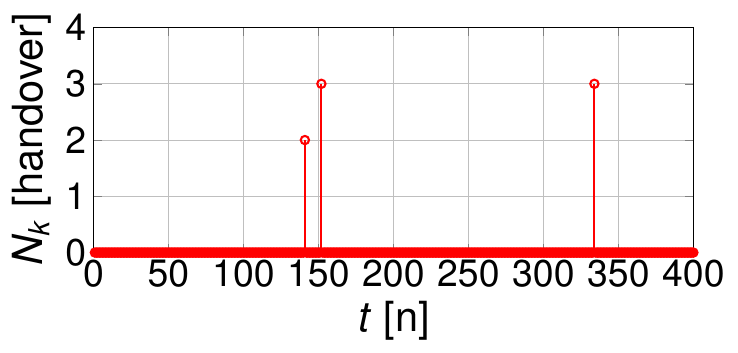}}
	\subfigure[$SE'$, \textit{nearOpt} (Alg. \ref{optimal})]{
		\label{SEvT.sub.3}
		\includegraphics[width=0.48\linewidth]{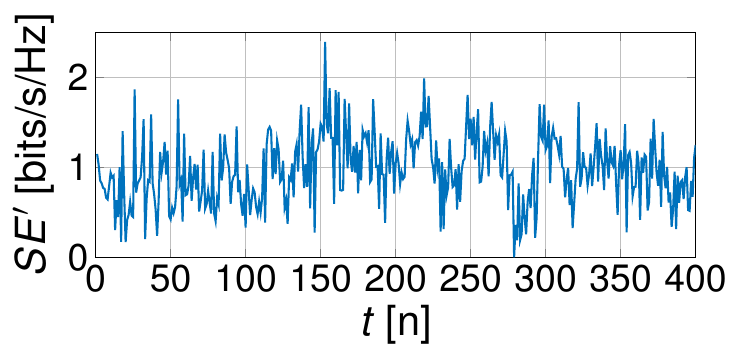}}
	\subfigure[$N_k$, \textit{nearOpt} (Alg. \ref{optimal})]{
		\label{SEvT.sub.7}
		\includegraphics[width=0.48\linewidth]{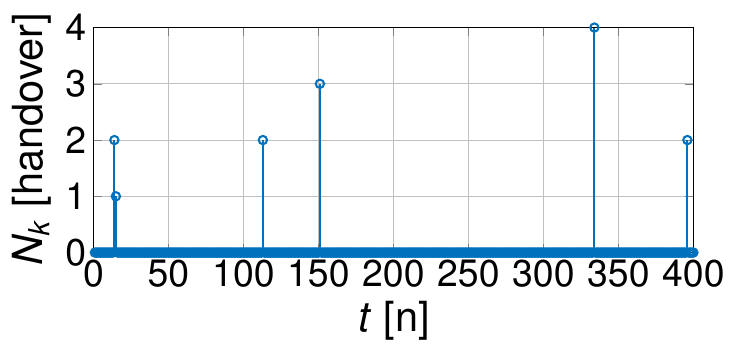}}
	\caption{Baseline SE $SE'$ and number of handovers $N_k$ of an example UE vs. time in CF-mMIMO with our proposed handover algorithms (PPP network, RWP UE mobility, \mbox{$v = 1$ m/s}).}
	\label{SEvTour}
\end{figure}

Let us now evaluate the performance of our proposed \textit{FairDiff} and \textit{nearOpt} handover algorithms under the same network in Sec. \ref{results:ref}. Fig. \ref{PPPuec.mobiSE} shows that \textit{FairDiff} significantly improves the SE under mobility compared to \mbox{\textit{always-handover}} and outperforms all three prior benchmarks for all UEs, and the difference to the scalable CF-mMIMO upper bound (i.e., baseline $SE'$ of \textit{always-handover}) is within 0.1 bits/s/Hz. Importantly, \textit{FairDiff} achieves similar throughput for the worst-served $95^\text{th}$ percentile UEs as the baseline SE of \textit{always-handover}. This is because \textit{FairDiff} \textit{both} significantly reduces the handover rate (\textit{cf.} Fig. \ref{PPPuec.HO}) \textit{and} achieves a comparably high baseline SE to the \textit{always-handover} case (\textit{cf.} Fig. \ref{PPPuec.baseSE}). \textit{FairDiff} achieves this by by differentiating the handover policy of different UEs based on the SNR quality of their serving AP sets, and thus suppresses the unnecessary handovers to reduce handover rate and encourage necessary ones to achieve high baseline SE. Consequently, Fig. \ref{SEvTb.sub.fairHO} shows that \textit{FairDiff} triggers precise handovers at the throughput degradation moment \mbox{$t=150n$}. Comparing Figs.~\ref{SEvTb.sub.fair} \& \ref{SEvTb.sub.fairHO} to \ref{SEvT.sub.3} \& \ref{SEvT.sub.7} further shows that \textit{FairDiff} exhibits a similarly good behavior as the near-optimal direct solution of our optimization problem \eqref{eq:optimizationSE} \textit{nearOpt}. This illustrates that our heuristic algorithm \textit{FairDiff} also achieves near-optimal performance in solving our optimization problem \eqref{eq:optimizationSE}. \looseness=-1

We emphasize that the good performance of our proposed \textit{FairDiff} and \textit{nearOpt} handover algorithms comes from directly maximizing the nett throughput under mobility (\textit{cf.} objective (\ref{objSE})). By doing so, our algorithms successfully assign necessary handovers to the UEs ``in need'' to update their AP serving set in a timely manner. Our algorithms thus achieve superior SE under mobility in Fig. \ref{PPPuec.mobiSE} compared to the three prior benchmarks that are designed under an incomplete objective, especially for the worst-served UEs. We emphasize that this is a particularly important result, since we restore the advantage of CF-mMIMO to deliver both high and \textit{uniform} throughput i.e., improve the performance of the poorly-served ``edge'' UEs, under mobility.

\subsubsection{In Realistic Urban Networks}
\label{otherReal}

\begin{figure}[!t] 
	\centering
	\includegraphics[width=1\columnwidth]{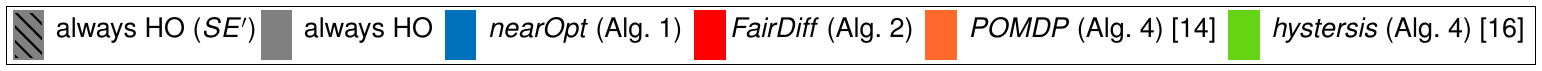}
	\subfigure[median SE]{
		\label{otherCitybar.SEm.uec}
		\includegraphics[width=1\linewidth]{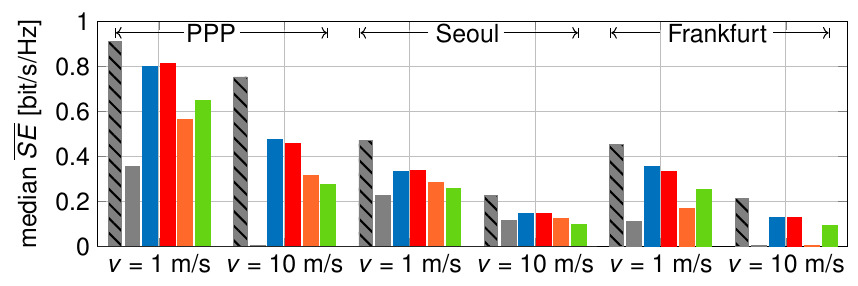}}
	\subfigure[$95^\text{th}$\%-ile SE]{
		\label{otherCitybar.SEw.uec}
		\includegraphics[width=1\linewidth]{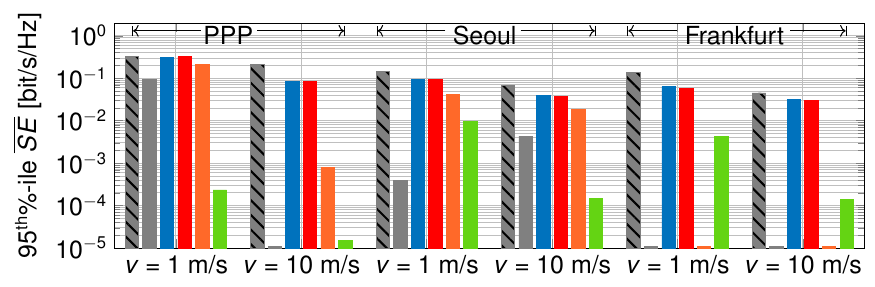}}
	\subfigure[handover rate]{
		\label{otherCitybar.ho.uec}
		\includegraphics[width=1\linewidth]{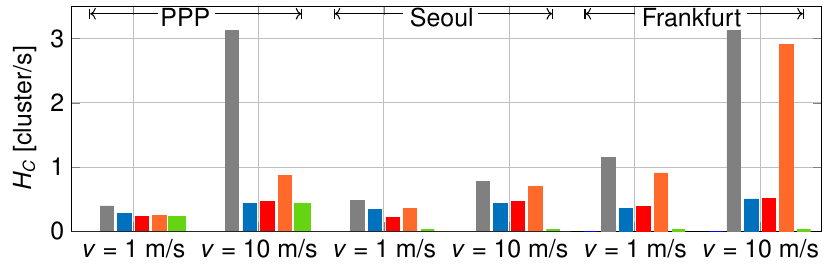}}
	\caption{Mobility-aware SE and handover rate with different network topologies, UE mobility patterns ($v=1$ m/s for pedestrian and $v=10$ m/s for vehicular mobility), and handover schemes.}
	\label{otherCitybar}
\end{figure}

We here extend the performance analysis of all considered handover schemes to different network  topologies and UE mobility patterns to evaluate whether the superior performance of our \textit{FairDiff} and \mbox{\textit{nearOpt}} handover schemes shown in an example PPP network in Sec.~\ref{myAlgoPPP} holds in general. \mbox{Fig. \ref{otherCitybar}} shows the SE under mobility given by (\ref{SEm}) of both the median and the \mbox{worst-served} ($95^\text{th}$ percentile) UEs and the handover rate given by (\ref{HOfinal}), for all considered networks. Since \textit{UPA} has very similar behavior as \textit{hysteresis}, we only show the performance of \textit{hysteresis} as a representative of both \textit{hysteresis} and \textit{UPA} in the rest of this paper. \looseness=-1

Overall, the SE performance trend of different handover schemes is similar in both PPP and urban networks. The SE under mobility decreases with the increases of UE speed due to stronger channel aging effects \cite{aging} and higher handover rates (\textit{cf. }Fig. \ref{otherCitybar.ho.uec}). Under the pedestrian UE speed $v=1$~m/s, \mbox{Figs. \ref{otherCitybar.SEm.uec}} and \ref{otherCitybar.SEw.uec} show that the SE of urban networks is in general lower than the PPP networks, due to higher handover rate (\textit{cf. }Fig. \ref{otherCitybar.ho.uec}). This is because the pedestrian walks in \mbox{Figs. \ref{city.sub.2}} and \ref{city.sub.3} tend to exhibit small but more frequent changes in direction than RWP in Fig. \ref{city.sub.1}. By contrast, vehicular tracks usually follow the road smoothly, leading to more directed paths and thus lower handover rates than RWP tracks. However, the SE of urban networks with vehicular mobility is still lower than that of the corresponding PPP networks, due to the \mbox{non-uniform} AP distribution and UE mobility patterns \cite{poster}. Furthermore, \mbox{Figs. \ref{otherCitybar.SEm.uec}} and \ref{otherCitybar.SEw.uec} show that the SE under mobility in Frankfurt is lower than that in Seoul under both pedestrian and vehicular mobility. This is because, as shown in \mbox{Figs. \ref{city.sub.3}} and \ref{city.sub.5}, there are many large building blocks in the Frankfurt area and the UEs tend to move around the buildings, and thus are more likely to go around CPU cluster edges over the whole path, incurring higher handover rates (\textit{cf. }Fig. \ref{otherCitybar.ho.uec}).

For the individual handover schemes, \mbox{Figs. \ref{otherCitybar.SEm.uec} and \ref{otherCitybar.SEw.uec}} show that the \mbox{\textit{always-handover}} case achieves significantly lower SE under mobility than the baseline SE $SE'$ for all considered UE speeds. Especially with the vehicular mobility \mbox{$v=10$} m/s, both the median and the \mbox{worst-performing} UEs suffer from link failure due to high handover rate (\textit{cf.} \mbox{Fig. \ref{otherCitybar.ho.uec}}), emphasizing the importance of a selective handover scheme in vehicular networks. Importantly, our \textit{FairDiff} scheme achieves consistently a superior SE under mobility than all prior benchmarks, which is comparable to the baseline SE upper bound of \textit{always-handover} and the near-optimal solution \textit{nearOpt}, in both PPP and urban networks. Fig. \ref{otherCitybar.ho.uec} shows that both \textit{FairDiff} and \textit{nearOpt} do so by obtaining similarly low handover rate in all considered networks, demonstrating their ability to perform precise handovers across different CF-mMIMO networks. By contrast, Fig. \ref{otherCitybar.ho.uec} shows that \textit{POMDP} barely reduces the handover rate compared to the \mbox{\textit{always-handover}} case and \textit{hysteresis} does not perform handover at all in the urban networks, resulting in their poor performance for both the median (\textit{cf. }\mbox{Fig. \ref{otherCitybar.SEm.uec}}) and the worst-served (\textit{cf. }\mbox{Fig. \ref{otherCitybar.SEw.uec}}) UEs. This again emphasizes that solely optimizing the baseline SE as \textit{POMDP} or solely reducing handover rate as \textit{hysteresis} cannot achieve uniformly high SE under mobility. In summary, both our \textit{nearOpt} and \textit{FairDiff} handover schemes achieve excellent mobile throughput performance and outperform all other prior benchmarks, regardless of network configuration. Our proposed schemes thus fulfill the promise of both \textit{high} and \textit{uniform} throughput of \mbox{CF-mMIMO} in both PPP and practical mobile networks.

\subsection{Analysis of the Proposed Algorithms}
\label{algoAna}


\subsubsection{Optimality Validation of \textit{nearOpt}}
\label{algoAna:nearOpt}

We here validate the near-optimal performance of our \textit{nearOpt} algorithm by evaluating the problem transformation given in Sec. \ref{optSolve}. 

\begin{figure}[!tb] 
	\centering{
		\includegraphics[width=0.95\linewidth]{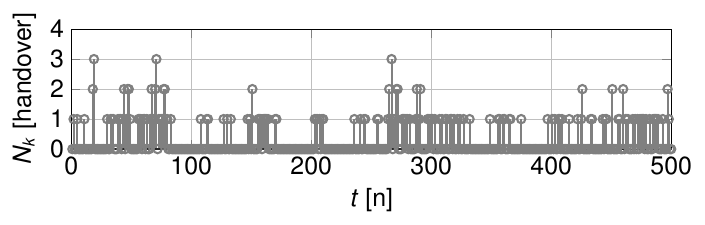}}
	\caption{Number of CPU cluster handovers $N_k$ of an example UE vs. time in UE-centric CF-mMIMO with the \textit{always-handover} case (PPP network, RWP UE mobility, \mbox{$v = 10$ m/s}). \looseness=-1}
\label{HOv10}
\end{figure}

\textbf{Validation of Transformation 2}: Fig.~\ref{HOv10} presents the number of CPU cluster handovers $N_k$ of an example UE in the \textit{always-handover} case versus time over its entire mobility path of $T=500$ communication blocks, in the PPP network with $v=10$ m/s. As Fig. \ref{otherCitybar.ho.uec} shows, this case represents the highest handover frequency among all considered scenarios and the typical urban UE speed. Fig. \ref{HOv10} shows that in around 90\% of the handover moments, the number of CPU cluster handover is at most 1. This confirms that the number of CPU cluster handovers $N_k[n] = 1$ is the most common case under mobility, confirming our Transformation 2.

\begin{figure}[!tb] 
	\centering{
		\includegraphics[width=0.95\linewidth]{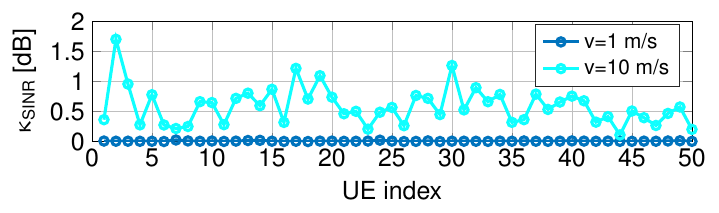}}
	\caption{SINR difference of all UEs in an example block in UE-centric CF-mMIMO (PPP network, RWP UE mobility). \looseness=-1}
\label{FigsinrT}
\end{figure}

\textbf{Validation of Transformation 3}: We observe in Fig. \ref{FigsinrT} the SINR difference of all $K=50$ UEs between the beginning and the end of the data transmission block, i.e. $\kappa_\mathrm{SINR} = \text{SINR}_k[\zeta] - \text{SINR}_k[t_0+n \tau_c]$, for an example communication block $n$. Under channel aging, the SINR monotonically decreases over the data transmission segment \cite{aging}, and therefore $\kappa_\mathrm{SINR}$ is the largest SINR gap within a block. Fig. \ref{FigsinrT} shows that for all UEs in the PPP network, $\kappa_\mathrm{SINR}$ is under 2 dB for both considered UE speeds. This confirms that the SINR within a communication block is similar, and thus can be approximated by the SINR at the beginning of the data transmission block, validating Transformation 3.

\begin{figure}[!tb] 
	\centering{
		\includegraphics[width=0.95\linewidth]{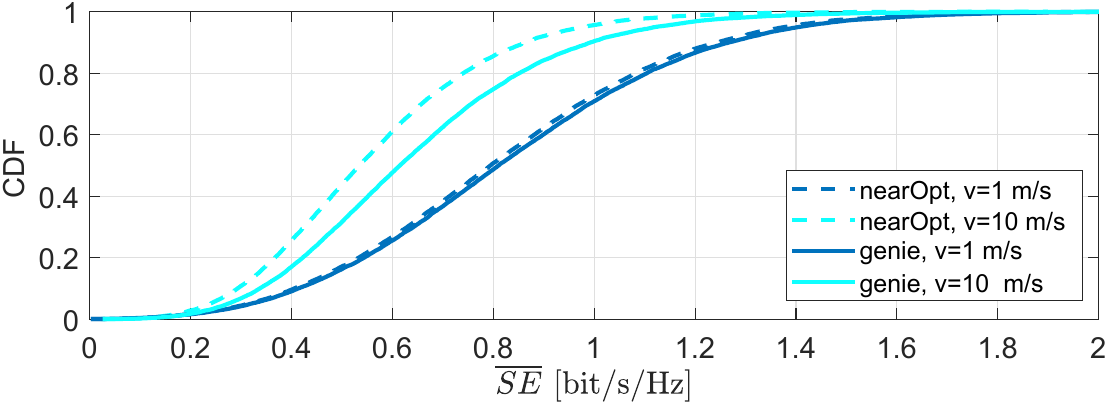}}
	\caption{Mobility-aware SE of \textit{nearOpt} and a genie-aided benchmark (PPP network, RWP UE mobility). \looseness=-1}
\label{genie}
\end{figure}

\textbf{Genie-Aided Benchmark}: We validate the optimality of our \textit{nearOpt} algorithm by benchmarking it against a genie-aided scheme that has access to the exact throughput model in Sec. \ref{throughput}. At each block $n$, this benchmark evaluates~\eqref{SEm} for both $x_k[n]=1$ and $x_k[n]=0$ using the exact SINR model in~\eqref{eqsi}, and selects the binary decision that yields the higher spectral efficiency, achieving per-block optimality. Since true global optimality over all blocks incurs prohibitive complexity and is unattainable by any benchmarks (\textit{cf. }Sec. \ref{optSolve}), this genie-aided benchmark is the strongest viable baseline. Fig.~\ref{genie} shows that \textit{nearOpt} achieves similar SE under mobility to the \mbox{genie-aided} benchmark at both $v=1$~m/s and $v=10$~m/s, confirming its near-optimality. The performance gap comes from the SINR approximations in Transformation 3 \& 4 and the numerical approximated solution of~\eqref{fprime} via Newton's method. The slightly larger gap at $v=10$~m/s than at $v=1$ m/s (0.05 bits/s/Hz versus 0.002 bits/s/Hz in the median SE) is due to more pronounced channel aging and \mbox{small-scale} fading at faster UE speeds, which increases the intra-block SINR variation (\textit{cf.}~Fig.~\ref{FigsinrT}) and the deviation between the exact SINR and its \mbox{large-scale} approximation in~\eqref{eqSimSINR}. We emphasize that the gap between \textit{nearOpt} and the genie-aided benchmark is not a consequence of suboptimal algorithm design, but reflects the inevitable cost of approximations that are necessary for realistic dense network deployment and are therefore inherent to any practical handover scheme. 


\subsubsection{Analysis of \textit{FairDiff}}
\label{algoAna:FairDiff}

We here analyse our \textit{FairDiff} algorithm in terms of the handover margin design, validity of threshold $\alpha$, and its robustness under AP selection schemes. \looseness=-1
 
\textbf{Margin Design}: We show in Table \ref{uecAB} the mobility-aware SE and handover rate of \textit{FairDiff} under different values of handover margins $\gamma_1$ and $\gamma_2$, in the PPP network with $v=1$~m/s (other networks exhibit the same performance trend). The throughput of \textit{FairDiff} increases with the decrease of $\gamma_1$ because a small $\gamma_1$ encourages the handover of the \mbox{worst-served} UEs. Furthermore, the throughput is insensitive to $\gamma_2$ because it is anyway hard to trigger handover for the \mbox{well-performing} UEs. Table \ref{uecAB} overall shows that over all considered handover margin values, \textit{FairDiff} achieves the best throughput performance when $\gamma_1=\gamma_2=1$ dB. We thus use this configuration in our performance analysis in \mbox{Sec. \ref{results}}. 


\begin{table}[!tb]
\centering
\caption{SE under Mobility [bit/s/Hz] and Handover Rate [cluster/s] for CF-mMIMO with \textit{FairDiff} in the PPP Network With $v=1$~m/s \\ ([median $SE_k$, $95^\text{th}$\%-ile $SE_k$], $H_c$)}
\footnotesize
\setlength{\tabcolsep}{2.5pt}
\renewcommand{\arraystretch}{0.95}

\begin{tabular}{|c|c|c|c|}
\hline
\backslashbox{$\gamma_1$}{$\gamma_2$}
& 1 dB & 4 dB & 10 dB \\ \hline

1 dB
& \makecell{\textbf{[0.81,0.33], 0.077}}
& \makecell{[0.80,0.32], 0.073}
& \makecell{[0.81,0.32], 0.073}
\\ \hline

4 dB
& \makecell{[0.81,0.30], 0.067}
& \makecell{[0.81,0.31], 0.060}
& \makecell{[0.81,0.31], 0.060}
\\ \hline

10 dB
& \makecell{[0.79,0.26], 0.063}
& \makecell{[0.77,0.25], 0.051}
& \makecell{[0.78,0.25], 0.051}
\\ \hline

\end{tabular}

\label{uecAB}
\end{table}

\begin{figure}[!tb] 
	\centering
	\subfigure[handover policy threshold $\alpha$]{
		\label{Pmean}
		\includegraphics[width=0.48\linewidth]{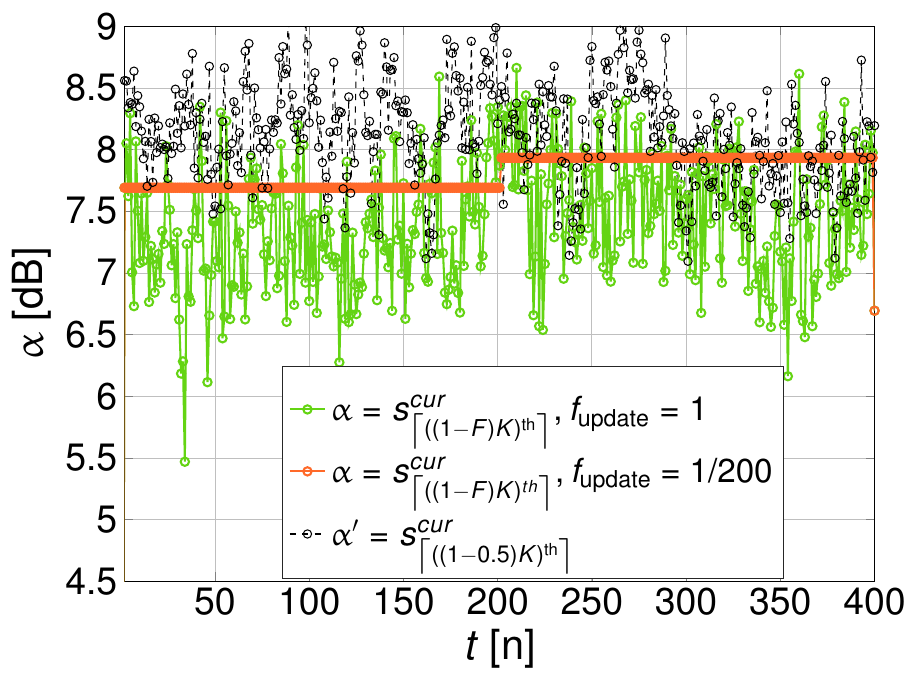}}
	\subfigure[spectral efficiency]{
		\label{Freq.PPP}
		\includegraphics[width=0.48\linewidth]{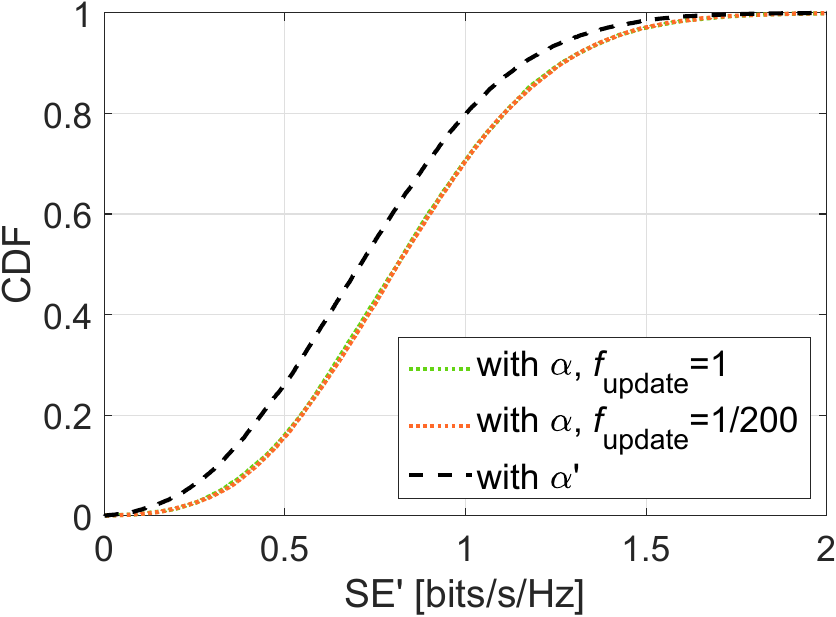}}
	\caption{Threshold $\alpha$ and mobility-aware spectral efficiency with different update frequencies for \textit{FairDiff} (PPP network, RWP UE mobility, $v = 1$ m/s).}
	\label{Freq}
\end{figure}

\textbf{Design of Threshold }$\bm{\alpha}$: We plot in Fig. \ref{Pmean} the threshold $\alpha$ value versus mobility time and in Fig. \ref{Freq.PPP} the SE of \textit{FairDiff} with different designs of $\alpha$, for the PPP network with \mbox{$v=1$~m/s}. To justify our design of $\alpha$ as a threshold obtained via Jain's fairness index $F$, in Fig. \ref{Freq} we also consider a variant design of the threshold $\alpha'= s^{cur}_{\left\lceil ((1-0.5)K)^{th} \right \rceil}$ which avoids computing $F$ and instead sets the threshold to the SNR of the median UE. This variant always considers half of the UEs as ``\mbox{fairly-served}'' and suppresses handover for them. Fig.~\ref{Pmean} shows that $\alpha'$ is higher than $\alpha$, thus incurring too many handovers and degrading the overall throughput in Fig.~\ref{Freq.PPP}. This is because $F$ in fact consistently exceeds 0.5, averaging 0.65 in our example, meaning that around 65\% of UEs should be classified as fairly-served and their handovers should be suppressed. Conversely, setting the threshold $\alpha$ corresponding to an incorrectly high $F$ would suppress necessary handovers and likewise degrade throughput; the \textit{hysteresis} algorithm that obtains low throughput in Fig.~\ref{PPPuec.mobiSE} is in fact an extreme case of \textit{FairDiff} with $F \equiv 1$ and $\alpha \equiv 0$. Fig.~\ref{Freq} thus shows that only the fairness index can accurately classify fairly- and \mbox{unfairly-served} UEs and properly differentiate the handover policy. \looseness=-1

Moreover, Fig. \ref{Pmean} shows that when updating $\alpha$ at every communication block, the change of $\alpha$ over time is within 2~dB. Namely, the threshold does not change rapidly over time and does not need to be \mbox{re-calculated} as frequently as in Step 1 of Alg. \ref{network}. This is because for a network with independent UE traffic, i.e., every UE can freely choose the available mobility track with equal chance, the performance of the whole network is stable. Namely, if one UE moves from a location with good channel conditions and serving AP sets to a location with bad conditions, leading to a rapid change of SNR, another UE typically moves from a bad condition to good, evening out the network-wide SNR change. Both the RWP and the realistic (\textit{cf.} Fig. \ref{city}) UE mobility patterns studied in this paper fit this assumption. We then confirm that the change of $\alpha$ in Fig.~\ref{Pmean} holds for all network configurations considered in this paper. Assuming $\alpha$ is updated every $b$ communication blocks, we define the update frequency as $f_{update}=1/b$. We show in \mbox{Fig.~\ref{Freq.PPP}} that there is a negligible difference in the SE under mobility of \textit{FairDiff} with different $f_{update}$. This indicates that the complexity of \textit{FairDiff} can be reduced with a small $f_{update}$ (\textit{cf. }Sec.~\ref{complexity}). \looseness=-1





\begin{figure}[!t] 
	\centering
	\includegraphics[width=1\columnwidth]{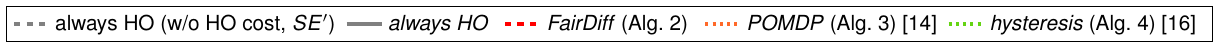}
    \subfigure[CoMP-JT AP selection]{
		\label{APselect.comp}
		\includegraphics[width=0.48\linewidth]{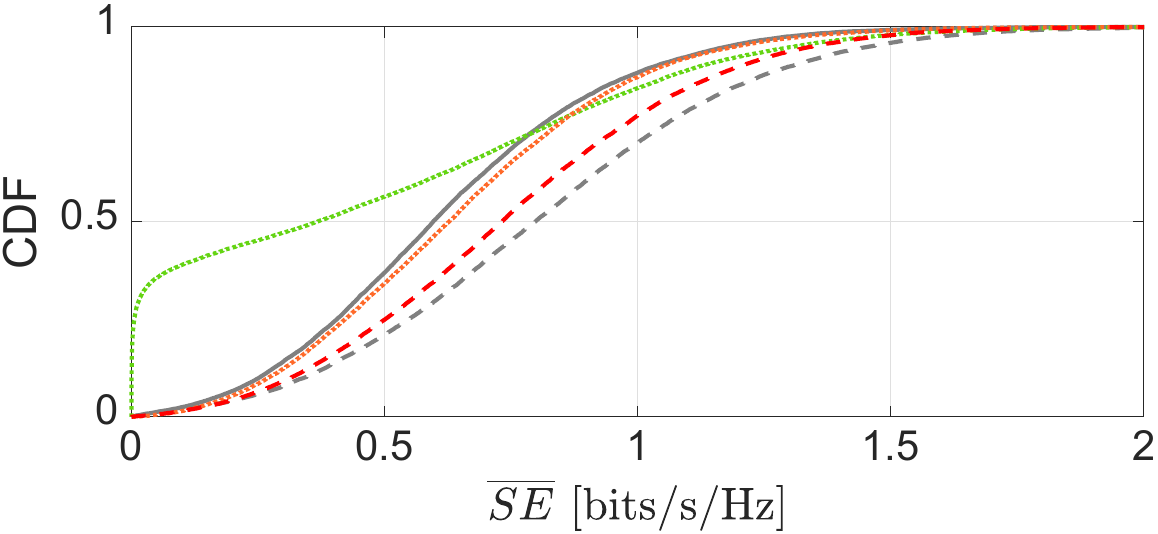}}
	\subfigure[PUC AP selection]{
		\label{APselect.puc}
		\includegraphics[width=0.48\linewidth]{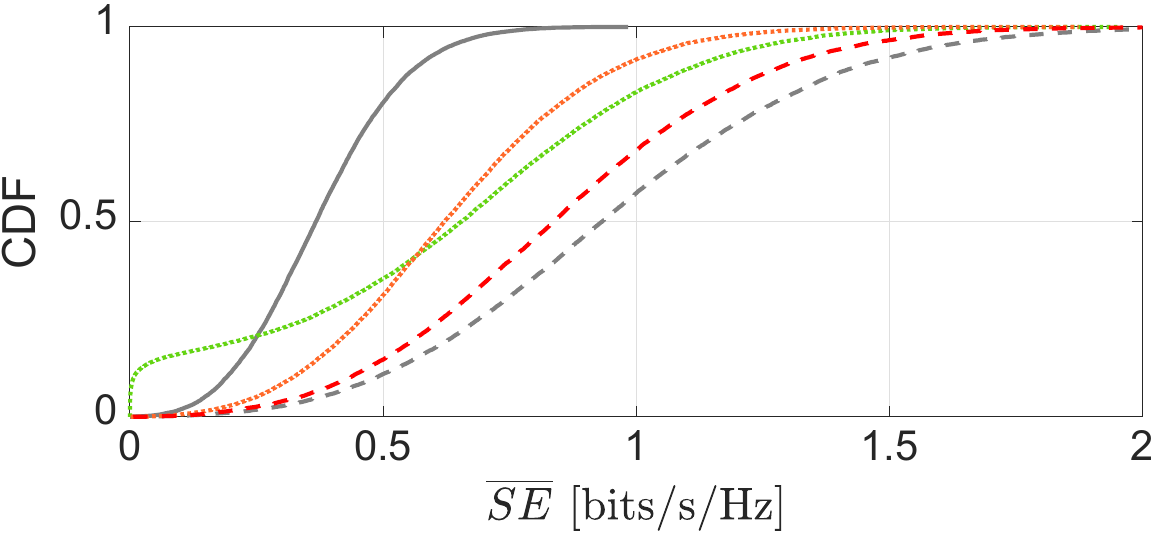}}
	\caption{Spectral efficiency of different AP selection and handover schemes (PPP network, RWP UE mobiltiy, $v = 1$ m/s).}
	\label{APselelct}
\end{figure}

\textbf{Performance With Different AP Selection}: Our \textit{nearOpt} solution can only applied to the CUC AP selection scheme in Sec. \ref{Arch} due to Transformation 1, while our \textit{FairDiff} algorithm is applicable to any AP selection scheme. We here compare the throughput of \textit{FairDiff} against all benchmarks under different AP selection schemes. Beside CUC in Fig.~\ref{PPPuec.mobiSE}, we also in Fig. \ref{APselelct} consider the coordinated multipoint joint transmission (\mbox{CoMP-JT}) AP selection where a given UE is served by one CPU cluster \cite{comp}, which is in fact CUC with $E=1$ and $Q=G$, and the pure \mbox{UE-centric} (PUC) AP selection where a UE is served by \mbox{$G$} APs with the best channel conditions \cite{scalableCF}. For a fair comparison, we set the average serving AP set size $G=34$ for all considered AP selection schemes. \mbox{Fig. \ref{APselelct}} shows that for all considered AP selection schemes, our \textit{FairDiff} algorithm significantly outperforms all prior benchmarks and achieves close throughput performance to the baseline SE of the \textit{always-handover} case, i.e., the upper bound for scalable \mbox{CF-mMIMO} under a given AP selection scheme. Fig. \ref{APselelct} thus confirms that the performance of our \textit{FairDiff} algorithm is robust across different AP selection schemes, while illustrating its better generalization ability than \textit{nearOpt}. \looseness=-1

\subsubsection{System Cost}
\label{complexity}

\begin{table}[!tb]
\vspace{-0.2cm}
\caption{Computational Complexity.}
\centering
\begin{tabular}{|c|c|}
\hline
Algorithm   & Complexity                          \\ \hline
\textit{nearOpt} (Alg. \ref{optimal}) & $\mathcal{O}\left(KM + \log(\log(1/\epsilon))\right)$ \\ \hline
\textit{FairDiff} (Alg. \ref{network})           & $\mathcal{O}(f_{update} (KM+K \log K)+KM)$                                    \\ \hline
\textit{POMDP} (Alg. \ref{pomdp} \cite{pomdp})            & $\mathcal{O}\left( 2^{2(G+1)} \right)$                                    \\ \hline
\textit{hysteresis} (Alg. \ref{hysteresis} \cite{beerten2025})           &   $\mathcal{O}\left(K M \right)$                                  \\ \hline
\textit{UPA} (Alg. \ref{drop} \cite{SoftHOcf})           & $\mathcal{O}\left(K M \right)$                                    \\ \hline
\end{tabular}
\label{bigO}
\end{table}

\begin{figure}[!tb] 
	\centering
	\includegraphics[width=1\columnwidth]{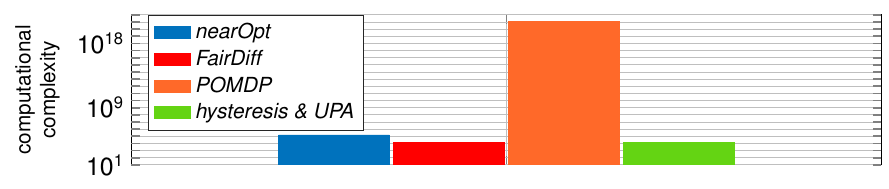}
	\caption{Computational complexity of different handover schemes, with \mbox{$M=308$}, $K=50$, and $G=34$.}
	\label{CC}
\end{figure}

We finally discuss the system cost of CF-mMIMO handover algorithms, in terms of signaling overhead and computational complexity. We firstly note that our \textit{FairDiff} and \textit{nearOpt} algorithms do not incur any extra signaling overhead compared to the benchmarks, which comes from transmitting the measured SNR for all UEs as the input of the handover algorithms and is the same for all considered algorithms. We furthermore compare the computational complexity of making handover decisions for all UEs at a decision moment for all considered handover algorithms, summarized in Table \ref{bigO}. For our \textit{nearOpt} algorithm, the computational complexity lies in solving the transcendental equation (\ref{fprime}), including the calculation of (\ref{eqA}) and (\ref{eqB}), i.e., obtaining the transformed SINR for the UE, and the complexity of the Newton method itself. The complexity is thus given by $\mathcal{O}\left(KM + \log(\log(1/\epsilon))\right)$ \cite{NewtonComplex}. Our \textit{FairDiff} scheme first requires the calculation of the total SNR per UE with a complexity of $\mathcal{O}(K M)$ in Steps 5 to 7, same as \textit{hysteresis} and \textit{UPA}. While the total SNR is the only calculation required for \textit{hysteresis} and \textit{UPA}, our \textit{FairDiff} additionally requires the calculation of the threshold $\alpha$, which is obtained via calculating the fairness index in (\ref{fi}) (Step 1 in Alg. \ref{network}) under the frequency $f_{update}$, with a complexity of $\mathcal{O}(KM)$ and sorting of the SNR of all UEs (Step 2 in Alg. \ref{network}) with a complexity of $\mathcal{O}(K\log K)$. The average computational complexity of $\alpha$ over the mobility period is then given by $\mathcal{O}(f_{update} (KM+K \log K))$. Finally, the computational complexity of \textit{POMDP} is dominated by the construction of the POMDP problem and given as $\mathcal{O}\left( 2^{2(G+1)} \right)$ in \cite{pomdp}, where $G$ is the average serving AP set size per UE. \looseness=-1


\mbox{Fig. \ref{CC}} gives the quantified computational complexity of all considered handover schemes for our considered network size (i.e., $M=308$, $K=50$, and $G=34$). The \textit{POMDP} scheme incurs significantly higher complexity than all other schemes due to the difficulty of solving the POMDP problem, and is thus not recommended for practical networks at all. Our \textit{nearOpt} solution achieves similarly good SE performance as our \textit{FairDiff} scheme (\textit{cf.} Secs. \ref{results:myAlgo}), but with an order of magnitude higher computational complexity, due to the solving of (\ref{fprime}). By contrast, our \textit{FairDiff} scheme entails similarly low complexity as \textit{hysteresis} and \textit{UPA} when we set \mbox{$f_{update}=1/200$} (justified via Sec. \ref{algoAna:FairDiff}), while achieving significantly superior throughput performance (\textit{cf.} Fig. \ref{otherCitybar}). \textit{FairDiff} is thus recommended for practical urban CF-mMIMO network deployments.

\vspace{-0.3cm}
\section{Conclusions}
\label{conclude}
We conducted a comprehensive performance analysis of mobile \mbox{CF-mMIMO} under realistic urban networks and proposed handover algorithms \textit{FairDiff} and \textit{nearOpt} to restore uniformly good performance of CF-mMIMO under practical mobility. We formulated and solved a novel optimization problem that maximizes the nett throughput under our comprehensive throughput model that considers both channel aging and handover cost. We derived a near-optimal solution \textit{nearOpt} by directly solving the relaxed optimization problem via Newton's method. We then proposed the low-complexity heuristic algorithm \textit{FairDiff} by introducing a handover policy threshold based on Jain's fairness index. Both our schemes achieve throughput close to the scalable \mbox{CF-mMIMO} upper bound and significantly outperform the prior benchmarks, especially for the \mbox{poorly-served} UEs, in both PPP distributed and realistic urban networks. Importantly, our \mbox{\textit{FairDiff}} scheme also entails low complexity. We thus proposed for the first time a handover scheme that delivers the original CF-mMIMO promise of uniformly good throughput in practical mobile networks. \looseness=-1


\vspace{-0.2cm}
\bibliographystyle{IEEEtran}
\bibliography{IEEEabrv, IEEEexampleMaster.bib}

@INPROCEEDINGS{xiao,
  title="{Mobility performance analysis of scalable cell-free massive MIMO}", 
    author={Yunlu Xiao and Petri M\"{a}h\"{o}nen and Ljiljana Simi\'{c}},
    booktitle={Proc. IEEE ICC}, 
    address = {Seoul},
    year={2022},
    volume={},
    number={}
}

@INPROCEEDINGS{wcnc,
  author={Yunlu Xiao and others},
  booktitle={Proc. IEEE WCNC}, 
  title="{A novel socially-differentiated handover scheme for UE-centric cell-free massive MIMO}", 
  address = {Dubai},
  year={2024},
  volume={},
  number={}
}

@INPROCEEDINGS{myGCpaper,
  title="{Mobility performance of scalable cell-free massive
MIMO under channel aging and handover}", 
    author={Yunlu Xiao and Ljiljana Simi\'{c}},
    booktitle={Proc. IEEE GLOBECOM}, 
    address = {Kuala Lumpur},
    year={2023},
    volume={},
    number={}
}

@INPROCEEDINGS{poster,
  author={Yunlu Xiao and Petri M\"{a}h\"{o}nen and Ljiljana Simi\'{c}},
  booktitle={Proc. IEEE INFOCOM}, 
  title="{Poster abstract: Performance of scalable cell-free
massive MIMO in practical network topologies}", 
  year={2023},
  address = {Hoboken, NJ, USA},
  volume={},
  number={}}

@article{MIMObook,
year = {2017},
volume = {11},
journal = {Foundations and Trends in Signal Processing},
title="{Massive MIMO networks: spectral, energy, and hardware efficiency}",
number = {3-4},
pages = {154-655},
author = {Emil Bj\"{o}rnson and others}
}

@ARTICLE{differdelay,  author={Lee, Won-Yeol and Akyildiz, Ian F.},  journal={IEEE Trans. Mobile Comput.},   title="{Spectrum-aware mobility management in cognitive radio cellular networks}",   year={2012}, month={Apr.},  volume={11},  number={4},  pages={529-542},  doi={10.1109/TMC.2011.69}}

@ARTICLE{tutorial,  author={Tabassum, Hina and others},  journal={IEEE Commun. Surveys Tuts.},   title="{Fundamentals of mobility-aware performance characterization of cellular networks: a tutorial}",   year={2019}, month={thirdquarter},  volume={21},  number={3},  pages={2288-2308},  doi={10.1109/COMST.2019.2907195}}

@ARTICLE{towards,  author={Lin, Xingqin and others},  journal={IEEE Trans. Wireless Commun.},   title="{Towards understanding the fundamentals of mobility in cellular networks}",   year={2013}, month={Apr.}, volume={12},  number={4},  pages={1686-1698},  doi={10.1109/TWC.2013.022113.120506}}

@ARTICLE{agingPredict,
  author={Truong, Kien T. and Heath, Robert W.},
  journal={J. Commun. Net.}, 
  title="{Effects of channel aging in massive MIMO systems}", 
  year={2013},
  month={Aug.},
  volume={15},
  number={4},
  pages={338-351},
  }

@ARTICLE{agingOptimize,
  author={Jiang, Wei and Schotten, Hans Dieter},
  journal={IEEE Commun. Lett.}, 
  title="{Impact of channel aging on zero-forcing precoding in cell-free massive MIMO systems}", 
  year={2021},
  month={Jul},
  volume={25},
  number={9},
  pages={3114-3118},
  }

@ARTICLE{pomdp,
  author={Ammar, Hussein A. and others},
  journal={IEEE Trans. Wireless Commun.}, 
  title="{Handoffs in user-centric cell-free MIMO networks: a POMDP framework}", 
  year={2024},
  month={Aug.},
  volume={23},
  pages={10319-10335},
  }

@ARTICLE{SoftHOcf,
  author = {Zaher, Mahmoud and others},
  journal={IEEE Trans. Wireless Commun.}, 
  title = "{Soft handover procedures in mmWave cell-free massive MIMO networks}",
  year={2024},
  month={Jun.},
  volume={23},
  number={6},
  pages={6124-6138}
  }

@ARTICLE{scalableaging,
  author={Zheng, Jiakang and others},
  journal={IEEE J. Sel. Areas Commun.}, 
  title="{Cell-free massive MIMO-OFDM for high-speed train communications}", 
  year={2022},
  month={Oct.},
  volume={40},
  number={10},
  pages={2823-2839},
  doi={10.1109/JSAC.2022.3196088}}

@ARTICLE{green,
  author={Mowla, Md Munjure and others},
  journal={IEEE Trans. Green Commun. Netw.}, 
  title="{A green communication model for 5G systems}", 
  year={2017},
  month={Sep.},
  volume={1},
  number={3},
  pages={264-280},
  doi={10.1109/TGCN.2017.2700855}}

@ARTICLE{aging,  author={Zheng, Jiakang and others},  journal={IEEE Trans. Wireless Commun.},   title="{Impact of channel aging on cell-free massive MIMO over spatially correlated channels}",   year={2021}, month={Oct.}, volume={20},  number={10},  pages={6451-6466},  doi={10.1109/TWC.2021.3074421}}

@INPROCEEDINGS{cfsim,
  author={Interdonato, Giovanni and Frenger, Pal and Larsson, Erik G.},
  booktitle={Proc. IEEE ICC}, 
  title="{Scalability aspects of cell-free massive MIMO}", 
  address = {Shanghai},
  year={2019},
  volume={},
  number={},
  doi={10.1109/ICC.2019.8761828}}

@ARTICLE{cfvs,
  author={Ngo, Hien Quoc and others},
  journal={IEEE Trans. Wireless Commun.}, 
  title="{Cell-free massive MIMO versus small cells}", 
  year={2017},
  month={Jan.},
  volume={16},
  number={3},
  pages={1834-1850},
  doi={10.1109/TWC.2017.2655515}}

@ARTICLE{scalableCF,
  author={Björnson, Emil and Sanguinetti, Luca},
  journal={IEEE Trans. Commun}, 
  title="{Scalable cell-free massive MIMO systems}", 
  year={2020},
  month={Jul.},
  volume={68},
  number={7},
  pages={4247-4261}}

@ARTICLE{NewtonComplex,
  author={Polyak, Roman A},
  journal={Pure \& Applied Functional Analysis}, 
  title="{Complexity of the regularized Newton's method}", 
  year={2018},
  month={Jun.},
  volume={3},
  number={2},
  pages={327-347}}

@ARTICLE{comp,  author={Chen, Shuyi and others},  journal={IEEE Network},   title="{Performance analysis of downlink coordinated multipoint joint transmission in ultra-dense networks}",   year={2017}, month={Aug.}, volume={31},  number={5},  pages={106-114},  doi={10.1109/MNET.2017.1600192}}

@techreport{3gppTlat,
 author = {3GPP},
 institution = {{3rd Generation Partnership Project (3GPP)}},
 note = {Version 18.1.0},
 number = {38.133},
 title = "{NR; Requirements for support of radio resource management}",
 type = {Technical Specification (TS)},
 year = {2023}
}

@techreport{3gpp211,
 author = {3GPP},
 institution = {{3rd Generation Partnership Project (3GPP)}},
 note = {Version 19.1.0},
 number = {38.211},
 title = "{5G; NR; Physical channels and modulation}",
 type = {Technical Specification (TS)},
 year = {2025}
}

@techreport{3gpp331,
 author = {3GPP},
 institution = {{3rd Generation Partnership Project (3GPP)}},
 note = {Version 18.6.0},
 number = {38.331},
 title = "{5G; NR; Radio Resource Control (RRC); Protocol specification}",
 type = {Technical Specification (TS)},
 year = {2025}
}

@techreport{3gpp133,
 author = {3GPP},
 institution = {{3rd Generation Partnership Project (3GPP)}},
 note = {Version 19.2.0},
 number = {38.133},
 title = "{5G; NR; Requirements for support of radio resource management}",
 type = {Technical Specification (TS)},
 year = {2025}
}

@ARTICLE{HOns3,
  author={Karmakar, Raja and others},
  journal={IEEE Trans. Mobile Comput.}, 
  title="{Mobility management in 5G and beyond: a novel smart handover with adaptive time-to-trigger and hysteresis margin}", 
  year={2022},
  month={Oct.},
  volume={22},
  number={10},
  pages={1-16},
  doi={10.1109/TMC.2022.3188212}}

@ARTICLE{MAcf,
  author={Chen, Shuaifei and others},
  journal={China Commun.}, 
  title="{Wireless powered IoE for 6G: Massive access meets scalable cell-free massive MIMO}", 
  year={2020},
  volume={17},
  number={12},
  pages={92-109},
  doi={10.23919/JCC.2020.12.007}}

@ARTICLE{beerten2025,
  author={Beerten, Robbert and others},
  journal={IEEE Open J. Commun. Soc.}, 
  title="{Mobile cell-free massive MIMO: a practical O-RAN-based approach}", 
  year={2025},
  month={Jan.},
  volume={6},
  number={},
  pages={593-610}
  }

@INPROCEEDINGS{shortFI,
  author={Deng, Jing and Han, Yunghsiang S. and Liang, Ben},
  booktitle={Proc. IEEE GLOBECOM}, 
  title="{Fairness index based on variational distance}", 
  year={2009},
  address={Honolulu}}

@ARTICLE{optimalFI,
  author={Sediq, Akram Bin and others},
  journal="{IEEE Trans. Wireless Commun.}", 
  title="{Optimal tradeoff between sum-rate efficiency and Jain's fairness index in resource allocation}", 
  year={2013},
  month={Jul.},
  volume={12},
  number={7},
  pages={3496-3509}
}

@article{viswalk,
title = "{Modeling pedestrian queuing using micro-simulation}",
journal = {Transportation Research Part A: Policy and Practice},
volume = {49},
pages = {232-240},
year = {2013},
month={Mar.},
author = {Inhi Kim and Ronald Galiza and Luis Ferreira},
}

@INPROCEEDINGS{mywcnc25,
  author={Xiao, Yunlu and Simi\'{c}, Ljiljana},
  booktitle={Proc. IEEE WCNC}, 
  title="{Performance of cell-free massive MIMO in realistic urban propagation environments}", 
  year={2025},
  address = {Milan}
  }

@electronic{shp,
  title         = "{OpenStreetMap data extracts}",
  url           = {https://download.geofabrik.de/},
  year          = {2025}
}

@ARTICLE{drlJP,
  author={Tsukamoto, Yu and others},
  journal={IEEE Open J. Commun. Soc.}, 
  title="{Scalable AP clustering with deep reinforcement learning for cell-free massive MIMO}", 
  year={2025},
  month={Feb.},
  volume={6},
  number={},
  pages={1552-1567}
}

@ARTICLE{DRLmaxMinRate,
  author={Banerjee, Bitan and others},
  journal={IEEE Trans. Mach. Learn. Commun. Netw.}, 
  title="{Access point clustering in cell-free massive MIMO using conventional and federated multi-agent reinforcement learning}", 
  year={2023},
  month={Jun.},
  volume={1},
  number={},
  pages={107-123}
  }

@ARTICLE{DRLmaxRate,
  author={Gao, Zhichao and others},
  journal={IEEE Commun. Lett.}, 
  title="{DRL-based AP selection in downlink cell-free massive MIMO network with pilot contamination}", 
  year={2024},
  month={Jun.},
  volume={28},
  number={6},
  pages={1432-1436}
}

@ARTICLE{DRLmobi,
  author={Prado, Anna and others},
  journal={IEEE J. Sel. Areas Commun.}, 
  title="{Enabling proportionally-fair mobility management with reinforcement learning in 5G networks}", 
  year={2023},
  month={Jun.},
  volume={41},
  number={6},
  pages={1845-1858}
}

@ARTICLE{GMM,
  author={Biswas, Pialy and others},
  journal={IEEE Trans. Mach. Learn. Commun. Netw.}, 
  title="{Optimal access point centric clustering for cell-free massive MIMO using Gaussian mixture model clustering}", 
  year={2024},
  month={May.},
  volume={2},
  number={},
  pages={675-687}
  }

@ARTICLE{ORAN2025,
  author={Cao, Yang and others},
  journal={IEEE J. Sel. Areas Commun.}, 
  title="{Implementation of a cell-free RAN system with distributed cooperative transceivers under ORAN architecture}", 
  year={2025},
  month={Mar.},
  volume={43},
  number={3},
  pages={765-779}
  }

@ARTICLE{LiORAN,
  author={Li, Feiyang and others},
  journal={IEEE Trans. Mobile Comput.}, 
  title="{Multiple CPUs cooperation for CF massive MIMO with mmWave fronthaul and backhaul}", 
  year={2025},
  month={Mar.},
  volume={25},
  number={3},
  pages={1-15}
  }

@ARTICLE{mobilityaware,
  author={Arshad, Rabe and others},
  journal={IEEE Commun. Lett.}, 
  title="{Mobility-aware user association in uplink cellular networks}", 
  year={2017},
  month={Nov.},
  volume={21},
  number={11},
  pages={2452-2455},
  doi={10.1109/LCOMM.2017.2729560}}

@techreport{3gpp300,
 author = {3GPP},
 institution = {{3rd Generation Partnership Project (3GPP)}},
 note = {Version 17.0.0},
 number = {38.300},
 title = "{5G; NR; NR and NG-RAN Overall description; Stage-2}",
 type = {Technical Specification (TS)},
 year = {2022}
}

@book{Newton,
  title="{Newton methods for nonlinear problems}",
  author={Deuflhard, Peter},
  year={2011},
  publisher={Springer}
}

@INPROCEEDINGS{vissim,
  author={Chen, Meiyan},
  booktitle={Proc. IEEE ICITBS}, 
  title={Modeling and Simulation Analysis of Road Network Based on VISSIM}, 
  year={2019},
  address={Changsha},
  volume={},
  number={},
  pages={32-35}
  }

@ARTICLE{CFmag2026,
  author={Buzzi, Stefano and others},
  journal={IEEE Commun. Mag.}, 
  title="{Why user-centric cell-free distributed MIMO systems will be the disruptive 6G technology}", 
  year={2026},
  month={Jan.},
  volume={},
  number={},
  pages={1-7},
  }

@INPROCEEDINGS{CFfirst,
  author={Ngo, Hien Quoc and others},
  booktitle={Proc. IEEE SPAWC}, 
  title="{Cell-free massive MIMO: uniformly great service for everyone}", 
  year={2015},
  address={Stockholm},
  volume={},
  number={}
  }

@INPROCEEDINGS{Rpilot,
  author={Parida, Priyabrata and Dhillon, Harpreet S.},
  booktitle={Proc. IEEE GLOBECOM}, 
  title="{Random sequential adsorption-based pilot assignment for distributed massive MIMO systems}", 
  year={2019},
  address={Waikoloa},
  volume={},
  number={}
  }

\end{document}